\documentclass[oldversion]{aa}  

\usepackage{amsmath}
\usepackage{graphicx}
\usepackage{txfonts}
\usepackage{natbib}

\bibpunct{(}{)}{;}{a}{}{,}

\newcommand{\sigmalos}{\sigma_{\rm los}}
\newcommand{\sigmapart}{\sigma_{\rm part}}
\newcommand{\sigmaorb}{\sigma_{\rm orb}}

\newcommand{\halflight}{R_{\rm hl}}
\newcommand{\halfmass}{R_{\rm hm}}
\newcommand{\rlag}{R_f}

\newcommand{\rsun}{~{\rm R}_\odot}

\newcommand{\binfrac}{F_{\rm M}}

\newcommand{\amin}{a_{\rm min}}
\newcommand{\amax}{a_{\rm max}}
\newcommand{\aeff}{a_{\rm eff}}

\newcommand{\pmin}{P_{\rm min}}
\newcommand{\pmax}{P_{\rm max}}

\newcommand{\mmin}{M_{\rm min}}
\newcommand{\mmax}{M_{\rm max}}
\newcommand{\mcl}{M_{\rm cl}}
\newcommand{\mdyn}{M_{\rm dyn}}

\newcommand{\msun}{~{\rm M}_\odot}
\newcommand{\qeff}{q_{\rm eff}}

\newcommand{\kms}{km\,s$^{-1}$}

\newcommand{\opik}{\"{O}pik}
\newcommand{\diff}{{\rm d}}

\begin{document}
   \title{The effect of binaries on the dynamical mass determination of star clusters}

   \author{M.B.N. Kouwenhoven
          \inst{1}
          \and
          R. de Grijs\inst{1.2}
          }

   \offprints{M.B.N. Kouwenhoven}

   \institute{Department of Physics and Astronomy, 
      University of Sheffield,
      Hicks Building, Hounsfield Road,
      Sheffield S3~7RH, United Kingdom\\
      \email{t.kouwenhoven@sheffield.ac.uk, r.degrijs@sheffield.ac.uk}
      \and 
      National Astronomical Observatories, Chinese Academy of Sciences,
      20A~Datun Road, Chaoyang District, Beijing 100012, P.~R.~China}

   \date{Received ---; accepted ---}

 

\abstract{

The total mass of distant star clusters is often derived from the virial theorem, using line-of-sight velocity dispersion measurements and half-light radii. Although most stars form in binary systems, this is mostly ignored when interpreting the observations. The components of binary stars exhibit orbital motion, which may increase the measured velocity dispersion, and may therefore result in a dynamical mass overestimation. In this paper we quantify the effect of neglecting the binary population on the derivation of the dynamical mass of a star cluster. We simulate star clusters numerically, and study the dependence of the derived dynamical mass on the properties of the binary population. We find that the presence of binaries plays a crucial role for very sparse clusters with a stellar density comparable to that of the field star population ($\sim 0.1$~stars\,pc$^{-3}$), as the velocity dispersion is fully dominated by the binary orbital motion. For such clusters, the dynamical mass may overestimate the true mass by up to an order of magnitude. For very dense clusters ($\ga 10^7$~stars\,pc$^{-3}$), binaries do not affect the dynamical mass estimation significantly. For clusters of intermediate density ($0.1 - 10^7$~stars\,pc$^{-3}$), the dynamical mass can be overestimated by $10-100\%$, depending on the properties of the binary population.

}

   \maketitle
%

\section{Introduction}

Observations have shown that the majority of the field stars are part of a binary or multiple system \citep[e.g.,][]{duquennoy1991}. Moreover, both observations and numerical simulations have indicated that this property is primordial: the vast majority of stars are formed in binary or multiple systems \citep[e.g.,][and references therein]{mathieu1994,mason1998,goodwinkroupa2005,kobulnicky2007,kouwenhoven_adonis,kouwenhoven_recovery}. Although this has been known for more than a decade, binaries are often not properly taken into account when analysing integrated spectral line data of young star clusters.

Young star clusters, with typical masses of $\mcl = 10^{3-6}\msun$ indicate recent or ongoing violent star formation. Their formation is often triggered by mergers and close encounters between galaxies. Only a fraction of these young massive star clusters evolve into old globular clusters, while a substantial fraction ($\sim 60-90\%$) may dissolve into the field star population within about 30~Myr \citep[see, e.g.,][for a review]{degrijs2007review}. In order to study the formation and fate of these star clusters, it is necessary to obtain good estimates of their total mass, stellar content, dynamics, and binary population. In this paper we focus on the derivation of the total mass and the properties of the binary population in particular.

There are two straightforward methods to determine the total mass of a star cluster. The first is based on a derivation from the luminosity of the cluster. This photometric mass determination \citep[e.g.,][]{mengel2005,ma2006} is independent of the assumptions about the properties of the binary population, as the total mass is derived from the integrated luminosity, to which each star (whether single, primary, or companion star) contributes. This method requires an {\it a priori} knowledge of the stellar mass distribution $f_{\rm M}(M)$ in the cluster (and hence of the mass-to-light ratio), accurate estimates of the age, distance, metallicity and interstellar extinction. The second method is based on the virial theorem: the dynamical mass, $\mdyn$, is derived from the (projected) half-light radius $\halflight$ and the line-of-sight velocity dispersion, $\sigmalos$. The half-light radius is often assumed to be equal to the half-mass radius $\halfmass$, i.e., no mass segregation is assumed to be present \citep[see, however,][]{boily2005,fleck2006}. An estimate of $\mdyn$ can be obtained using the equation derived by \cite{spitzer1987}:
\begin{equation} \label{equation:spitzer}
  \mdyn = \eta \, \frac{\halflight  \sigmalos^2}{G} \,,
\end{equation}
where $G$ is the gravitational constant, and $\eta$ is a dimensionless proportionality constant. Spitzer's equation is valid under the following assumptions: (i) the cluster follows a Plummer density model, (ii) all stars are equal-mass stars, (iii) no binary or multiple stars are present and (iv) the cluster is in virial equilibrium. An estimate of the accuracy of the derived masses can be obtained by comparing photometric and dynamical masses \citep[e.g.,][]{mandushev1991,smith2001,larsen2004,maraston2004,degrijs2005,bastian2006,larsen2007}.

The Plummer model \citep{plummer1911} assumed in Eq.~(\ref{equation:spitzer}) describes the structure and dynamics of mature star clusters with reasonable accuracy. It has been in use for a long time because of its mathematical simplicity. \cite{king1962,king1966} developed a set of models, nowadays known as King models, that provide a more accurate description for globular clusters \citep[see, e.g.,][]{meylan1997}. Although Eq.~(\ref{equation:spitzer}) is derived for the Plummer model, it is also a good approximation for King models, with a slightly adjusted proportionality constant $\eta$ \citep[see, e.g.,][]{fleck2006}. Very young ($\la 1$~Myr) clusters, however, often exhibit an irregular or flocculent structure \citep[e.g.,][]{elmegreen2000}, so that Eq.~(\ref{equation:spitzer}) may not be a good approximation anymore. The equal-mass assumption is clearly inappropriate, although adopting a more realistic mass distribution affects the derived $\mdyn$ only mildly (see \S\,\ref{section:aperture}). The assumption of virial equilibrium is probably appropriate for older ($\ga 50$~Myr) clusters. It is, however, an incorrect assumption for most young ($\la 20$~Myr) clusters; a substantial fraction of these may suffer from infant mortality \citep[e.g.,][and references therein]{goodwinbastian2006,degrijs2007review}; we will briefly discuss this issue in \S\,\ref{section:virialratio}. In this paper, however, we focus primarily on the effects of assumption (iii),  the assumed absence of binary systems, and study how the parameter $\eta$ depends on the properties of the star cluster and the binary population.


In a cluster consisting of single stars, the velocity dispersion traces the motion of each particle (i.e., star) in the cluster potential. In a cluster with binary stars, on the other hand, we do not measure the motion of each particle (i.e., the binary centre-of-mass), but of the individual binary components. These have an additional velocity component due to their orbital motion, which may result in an overestimation of the dynamical cluster mass (e.g., \citealt{bosch2001}; \citealt{fleck2006}; \citealt{apai2007}).
%
For a very sparse, dissolving cluster ($\la 0.1$~stars\,pc$^{-3}$), the motion of the centre-of-mass of each binary in the cluster is much smaller than the orbital motion of the binary components. The spectral line width in such a cluster is thus dominated by orbital motion, and is not representative of the motion of the binaries in the cluster potential. Unless this effect is corrected for, the derived dynamical mass could be significantly overestimated. On the other hand, for a very dense cluster ($\ga 10^7$~stars\,pc$^{-3}$), the effect of binaries is almost negligible. Nevertheless, the presence of binaries always leads to a smaller value of $\eta$. If these binaries are not properly taken into account, the overestimation in the derived dynamical mass is given by $\mdyn/\mcl = 9.75\,\eta^{-1}$, where $\mcl$ is the true cluster mass. In our analysis we consider three types of clusters:
\begin{itemize}
\item {\em Particle-dominated clusters}. 
  The measured velocity dispersion is dominated by the 
  motion of the stars/binaries in the cluster potential. 
  Clusters with a low binary fraction or a large stellar density ($\ga 10^7$~stars\,pc$^{-3}$)
  are good examples. In this case Eq.~(1) 
  applies and $\eta\approx 9.75$.
\item {\em Intermediate clusters}. 
  Most realistic clusters are of this type. If the canonical value $\eta=9.75$ is 
  adopted, the derived value of $\mdyn$
  may mildly overestimate the true cluster mass, $\mcl$.
\item {\em Binary-dominated clusters}. 
  The measured velocity dispersion is dominated by 
  the orbital motion of the binaries, so that $\eta \ll 9.75$. Eq.~(1) may
  result in a significant overestimation of $\mdyn$ if the presence of binaries is ignored. 
  Binary-dominated 
  clusters generally have a low stellar density ($\sim 0.1$~stars\,pc$^{-3}$)
  and a high binary fraction.
\end{itemize}

The presence of binaries additionally affects the star cluster dynamics in the following way. Imagine a cluster consisting of $N$ stars. Suppose we now add companions to each of these stars. The number of particles (i.e., singles and binaries) is still $N$, but the mass of the cluster has increased, resulting in larger centre-of-mass velocities than in the case of a single-star cluster. We will return to this issue in \S\,\ref{section:comparisonissues}.

This paper is organised as follows. In \S\,\ref{section:method} we briefly describe our method and assumptions. We discuss the effect of varying the star cluster properties, such as the size, stellar density, mass, number of stars, virial ratio, stellar mass distribution, and the aperture size on the dynamical mass estimate in \S\,\ref{section:clusterproperties}. Subsequently, we study the effect of varying the binary population properties (binary fraction, mass ratio distribution, eccentricity distribution, and semi-major axis or period distribution) on the dynamical mass estimate in \S\,\ref{section:binarypopulationproperties}. By varying each of these properties we study their respective contribution to the value of $\eta$ in Eq.~(\ref{equation:spitzer}). In \S\,\ref{section:whencanbinariesbeignored} we describe, from a practical point of view, under which conditions binarity can be ignored, and under which conditions ignoring the binaries results in a significant overestimation of the dynamical mass. Finally, we summarise and discuss our results in \S\,\ref{section:conclusions}.


\section{Method and assumptions} \label{section:method}

We study the effect of binaries on the dynamical cluster mass determination using simulated clusters. In our numerical simulations we can determine $\mdyn$, $\sigmalos$, and $\halfmass$ for each cluster, allowing us to derive the true $\eta$ for clusters with different properties, and to study the error that is introduced in $\mdyn$ when binarity is ignored (i.e., if the canonical value $\eta=9.75$ is adopted).

\subsection{Model properties}

We simulate star clusters using the {\tt STARLAB} package \citep[see, e.g.,][]{ecology4}. We draw $N$ stars from a mass distribution $f_M(M)$ ($\mmin \leq M \leq \mmax$), such that the total mass of the cluster, after inclusion of the binary companions, is $\mcl=10^4\msun$ by default.  
We choose this value because it is a typical mass for open cluster-like objects, and because the effects of binarity on the dynamical mass are pronounced for such clusters. Dynamical masses for such low-mass clusters are derived in, for example, \cite{mandushev1991} and \cite{degrijs2008lowmass}. Note, however, that most well-studied clusters have {\em measured} dynamical masses of $10^{5-7}\msun$; this is mainly because of (1) their brightness, and (2) their large velocity dispersion. We will discuss the full range of cluster masses in \S~\ref{section:numberofparticles} and \S~\ref{section:whencanbinariesbeignored}.
We define the binary fraction as $\binfrac \equiv B/(S+B)$, where $S$ and $B$ are the number of single stars and binary systems in the cluster, respectively. The total number of ``particles'' is indicated by $N=S+B$ (note that this is not the number of individual stars $S+2B$). We ignore the presence of triple and higher-order systems.

A fraction $\binfrac$ of the stars are assigned a companion star. The mass ratio, semi-major axis (or, alternatively, the orbital period) and eccentricity of each binary are drawn from the respective probability distributions $f_q(q)$, $f_a(a)$ and $f_e(e)$. We assume random orientation for the binary star orbits. Each particle (i.e., single star or binary) is given a certain position and velocity according to the Plummer model \citep{plummer1911}, using the {\tt makeplummer} routine in the {\tt STARLAB} package. The cluster is scaled, such that it has a certain {\em projected} half-mass radius $\halfmass$. For Plummer models, the corresponding intrinsic half-mass radius is given by $R_{\rm hm,intr} = (2^{2/3}-1)^{-1/2}\halfmass$ \citep[e.g.,][]{heggiehut}. The average mass density within the half-mass radius is given by $(\tfrac{1}{2}\mcl)/(\tfrac{4}{3}\pi\halfmass^3) = \tfrac{3}{8}\pi^{-1} \mcl \halfmass^{-3}$. Similarly, the corresponding average stellar density is given by 
\begin{equation} \label{equation:density}
\langle \rho \rangle_{\rm hm} = \tfrac{3}{16}\,\pi^{-1}\, \langle M_T \rangle \,\mcl \,\halfmass^{-3} \,,
\end{equation}
where $\langle M_T \rangle$ is the average mass of a binary system.
Each cluster is assumed to be in virial equilibrium, and no mass segregation is assumed to be present. 

\subsection{Two canonical models}

In order to study the effect of each star cluster property separately, we perform our simulations with two different models, which we will refer to as (the simplified) model~S and (the more realistic) model~R. The default properties of the two models are listed in Table~\ref{table:twomodels}. Throughout this paper we vary each property of the binary population individually, keeping the other properties constant, in order to study the effect of each binary parameter on $\eta$ individually.

Model~S is a simplified star cluster model, consisting of equal-mass stars. If the binary fraction in Model~S is set to $\binfrac=0\%$, this model satisfies the assumptions of Eq.~(\ref{equation:spitzer}). We use this simplistic model, as changes of the cluster properties and the binary population have very pronounced effects on the derived $\sigmalos$ and $\eta$, thus allowing us to quantify the relations precisely. We assign a mass $M=1\msun$ to each star, and adopt a binary fraction of $\binfrac=100\%$. We assign to each binary a mass ratio $q\equiv M_2/M_1 = 1$, an eccentricity $e=0$, and a semi-major axis $a=10^3\rsun$ ($\approx 4.65$\,au).

Model~R, on the other hand, is more realistic, and is a good approximation for real star clusters. Its properties are identical to those of model~S, except for those mentioned below. Each star is assigned a mass which is drawn from the Kroupa initial mass function (IMF) \citep{kroupa2001}, in the mass range $0.02-20\msun$, given by
\begin{equation} \label{equation:kroupaimf}
  f_{\rm Kroupa}(M) = \frac{\diff N}{\diff M} \propto \left\{
  \begin{array}{llll}
    M^{-0.3}  & {\rm for \quad } 0.02 & \leq M/{\rm M}_\odot & < 0.08 \\
    M^{-1.3}  & {\rm for \quad } 0.08 & \leq M/{\rm M}_\odot & < 0.5   \\
    M^{-2.3}  & {\rm for \quad } 0.5  & \leq M/{\rm M}_\odot & \leq 20   \\
  \end{array}
  \right. .
\end{equation}
The Kroupa IMF is currently considered to be a good description of the stellar mass distribution. In our analysis we only draw stellar masses above the deuterium burning limit: $M=0.02\msun$, as lower-mass objects (such as planets) barely contribute to the dynamics of the cluster, the dynamical mass determination or the luminosity of the cluster. 
We adopt a maximum stellar mass of $20\msun$. Stars more massive than $20\msun$ are bright, but very rare. As we calculate the velocity dispersion directly from the line-of-sight velocities (i.e., we do not apply luminosity weighting), the results presented in the paper are practically independent of the choice for the the upper mass limit.
The binary fraction for model~R is 100\%. We adopt a semi-major axis distribution of the form $f_a(a)\propto a^{-1}$ ($10~\rsun \leq a \leq 0.02~\mbox{pc}$), a flat mass ratio distribution $f_q(q) = 1$, and a thermal eccentricity distribution $f_e(e)=2e$. The latter choices are motivated in  \S\,\ref{section:semimajoraxis}, \ref{section:massratio}, and~\ref{section:eccentricity}, respectively.

\begin{table}
  \caption{The default properties of the two models used in our analysis: the simple model~S (middle column) and the more realistic model~R (right-hand column). In our analysis we vary the properties of each model in order to find the effect of this change on the value of $\eta$. At the bottom of the table we list for each model the line-of-sight velocity dispersion $\sigmalos$ of the individual stars, $\sigmapart$ of the centre-of-mass motion of the binaries, and $\sigmaorb$ of solely the orbital motion of the binary components. Each value represents the width of the best-fitting Gaussian.
\label{table:twomodels}}
  \begin{tabular}{lll}
    \hline
    \hline
    Property         & Model S                 & Model R \\
    \hline
    Model            & Plummer                 & Plummer \\
    Proj. half-mass radius & $\halfmass=5$~pc        & $\halfmass=5$~pc\\
    Particles $N=S+B$& $N=5\,000$              & $N=18\,600$ \\
    Total mass       & $\mcl=10^4\msun$        & $\mcl=10^4\msun$ \\
    Mass segregation & No                      & No \\
    Virial equilibrium& Yes                    & Yes \\
    \hline 
    Primary mass     & $M_1 = 1\msun$            & $f_{\rm Kroupa}(M_1)$; $0.02-20\msun$ \\
    Binary fraction  & $\binfrac=100\%$        & $\binfrac=100\%$ \\
    Mass ratio       & $q=1$                   & $f_q(q) = 1$; $0<q\leq 1$ \\
    Eccentricity     & $e=0$                   & $f_e(e) = 2e$; $0\leq e < 1$ \\
    Orbital size     & $a=10^3 \rsun$          & $f_{\rm Opik}(a)$; $10\rsun-0.02$~pc \\
    Orientation      & Random                  & Random \\
    \hline
    $\sigmalos$  (\kms) & 7.10        & 1.20 \\
    $\sigmapart$ (\kms) & 0.91        & 0.91 \\
    $\sigmaorb$  (\kms) & 7.09        & 0.14 \\
    \hline
    \hline
  \end{tabular}
\end{table}

\subsection{The velocity dispersion}


\begin{figure}[!bt]
  \centering
  \begin{tabular}{c}
    \includegraphics[width=0.48\textwidth,height=!]{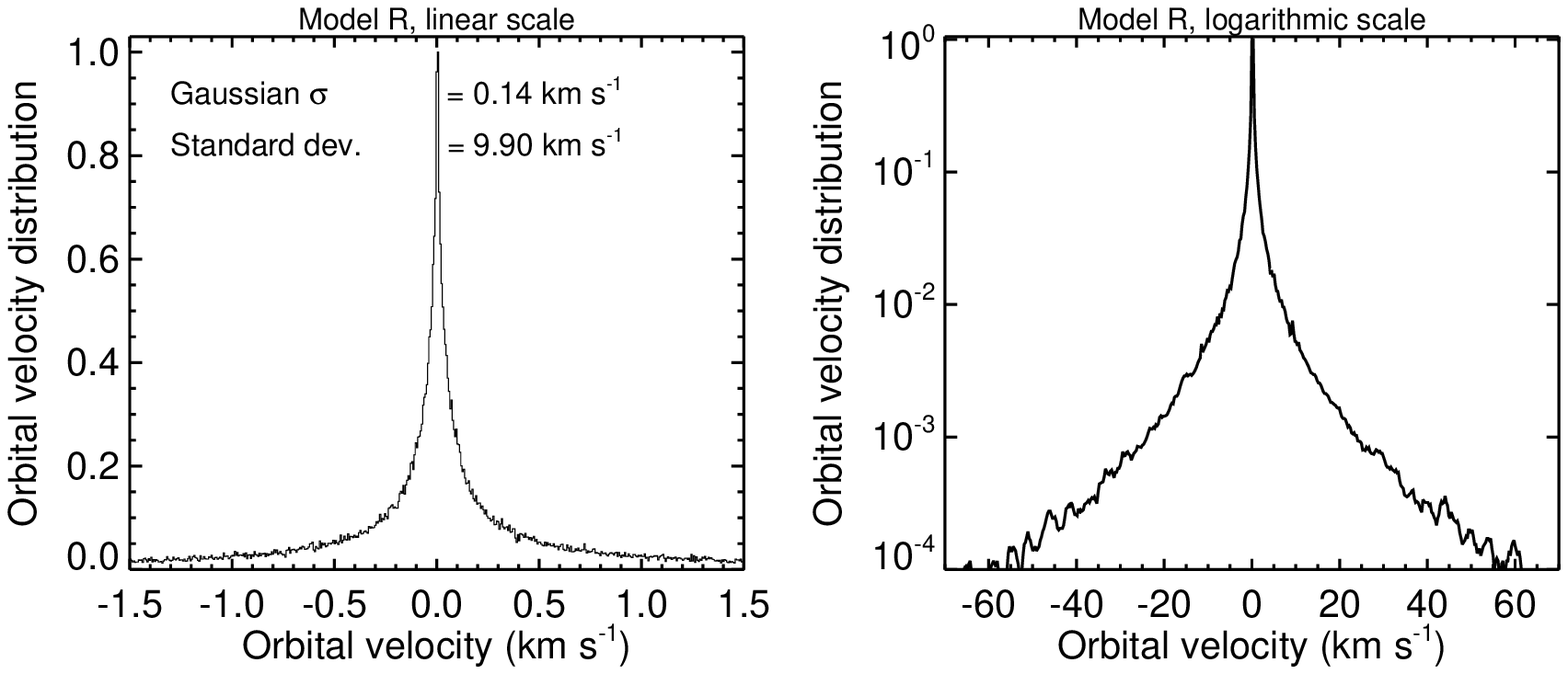}\\
    \includegraphics[width=0.48\textwidth,height=!]{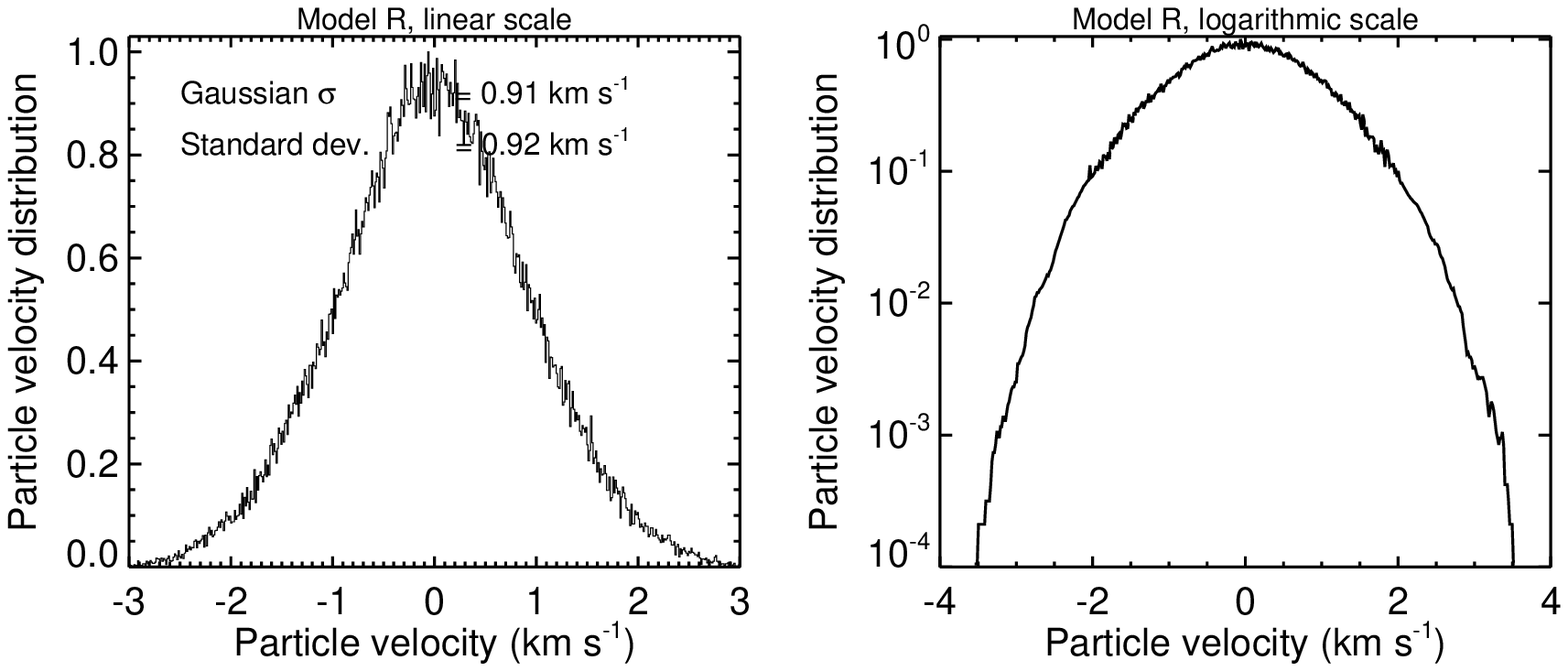}\\
    \includegraphics[width=0.48\textwidth,height=!]{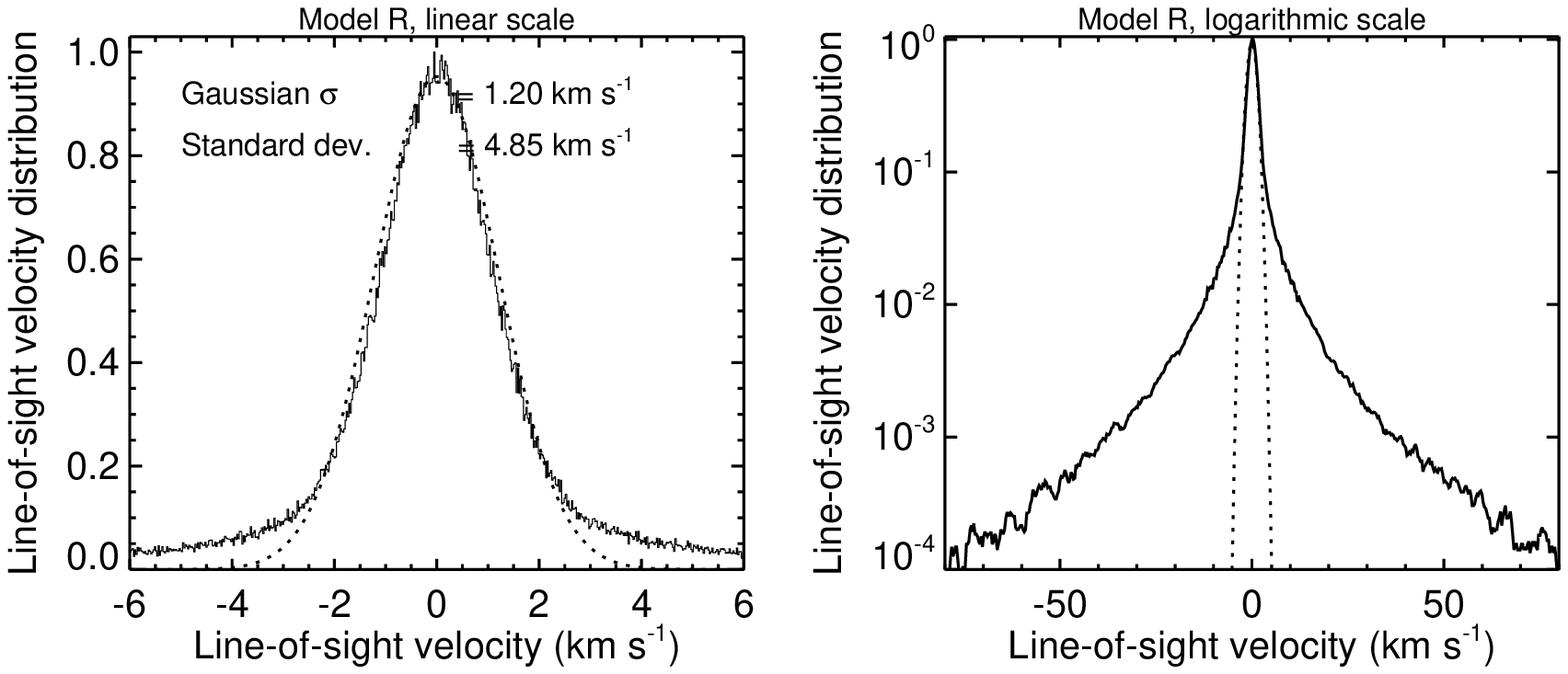}
  \end{tabular}
  \caption{The contribution of the (centre-of-mass) particle motion and the (binary) orbital motion to the measured velocity dispersion of a stellar population. The example shown here is for model~R with $\halfmass=5$~pc and $\mcl=10^4\msun$ and a binary fraction of 100\%. {\em Top panels:} the line-of-sight binary orbital velocity distribution $f_{v_{\rm orb}}(v_{\rm orb})$ for model~R, on a linear ({\em left}) and a logarithmic scale ({\em right}). Note that  $f_{v_{\rm orb}}(v_{\rm orb})$ is independent of the cluster size, mass, and binary fraction. {\em Middle panels:} Same for the line-of-sight centre-of-mass particle velocity distribution $f_{v_{\rm  part}}(v_{\rm part})$, i.e., only taking into account the motion of the centre-of-mass in the cluster. The particle velocity distribution is well fitted by a Gaussian distribution. Unlike the distribution in the top panels, this distribution depends strongly on the size and mass of the star cluster.  {\em Bottom panels:} the {\em measured} line-of-sight velocity dispersion of the binary components in the simulated cluster. This distribution is a combination of $f_{v_{\rm orb}}(v_{\rm orb})$ and $f_{v_{\rm  part}}(v_{\rm part})$. The best-fitting Gaussian is indicated with the dotted curve. The Gaussian function fits the central data well. The broad wings are not well fitted by the Gaussian; however, these are dominated by noise in real observations.
    \label{figure:vorbvpartvlos} }
\end{figure}


Direct measurements of the velocity dispersion of a star cluster can be done in three ways. Most commonly, $\sigmalos$ is determined from (i) the width of spectral lines from observations integrated over a large part of the cluster \citep[e.g.,][]{bastian2006,moll2007}. For nearby clusters the velocities of individual stars can be measured, both (ii) radial velocities \citep[e.g.,][]{reijns2006,apai2007} and (iii) proper motions  \citep[e.g.,][]{vanleeuwen2000,chen2007,stolte2007}. Due to the nature of the observations, the velocity dispersion obtained using techniques (i) and (ii) may be affected by the presence of binaries, while that obtained using technique (iii) is insensitive to binaries. Our paper thus applies to the spectral line and radial velocity studies, and not to proper motion studies.


In our simulations we obtain radial velocities $v_r$ for each star in our simulated cluster. By fitting Gaussian profiles we derive the projected velocity dispersion $\sigmalos$ from the projected radial velocity distribution $f_{v_r}(v_r)$ of the stars in the cluster. We consider only radial velocities in the range $\langle v_r \rangle - 3\,\mbox{rms}(v_r) \leq v_r \leq \langle v_r \rangle + 3\,\mbox{rms}(v_r)$, where $\langle v_r \rangle$ is the mean radial velocity and $\mbox{rms}(v_r)$ the corresponding root-mean-squared variation. Note that $\sim 99.5\%$ of the stars have line-of-sight velocities between these limits. We use this cut-off in $v_r$ for the following reasons. The distribution over velocity is generally not Gaussian, and has broad wings (see Fig.~\ref{figure:vorbvpartvlos}). The rare, extreme-velocity stars in these wings affect the determination of $\sigmalos$ significantly. For example, the (dotted) fit in the bottom panels of Fig.~\ref{figure:vorbvpartvlos} would be broader without the rejection of the 0.5\% of the stars with extreme velocities. A similar cut-off is automatically imposed during spectral line analysis of unresolved star clusters, where the low signal-to-noise broad wings remain undetected, and the best-fitting Gaussian is essentially determined using the brighter, central part of the spectral line.

The quantity $\sigmalos$ is affected by (i) the motion of the particles in the cluster potential, and (ii) the orbital motion of binary components about their centre-of-mass. We will refer to the orbital motion of the particles (i.e., the centres-of-mass) in a cluster as $\sigmapart$. We use $\sigmaorb$ to refer to the binary orbital motion, relative to their centres-of-mass. For the determination of the total cluster mass it is important to know whether one of these dominates. We refer to a cluster with $\sigmalos \approx \sigmapart$ as {\em particle-dominated}, and to a cluster with $\sigmalos \approx \sigmaorb$ as {\em binary-dominated}. For most clusters, however, both the particle motion and orbital motion are important; we refer to these as the {\em intermediate case}. Spitzer's equation is only applicable in the particle-dominated case, and results in an overestimation of $\mdyn$ in the intermediate and the binary-dominated cases.

For a binary population similar to that of model~R (Table~\ref{table:twomodels}), the velocity dispersion resulting from orbital motion is $\sigmaorb \approx 0.14~\mbox{km\,s}^{-1}$  Here we have included all binaries with equal weight, irrespective of their mass. This value is (by definition) independent of the binary fraction $\binfrac$. The weight given to $\sigmaorb$ relative to $\sigmapart$, when inferring $\sigmalos$, however, depends on $\binfrac$. Note that the above value of $\sigmaorb$ is obtained using all stars in the cluster, although in reality the velocity dispersion is obtained for a very specific subset of stars. The latter selection effect may change the effective value of $\sigmaorb$ significantly (see, e.g., \S\,\ref{section:massratio} and \S\,\ref{section:eccentricity}).


\subsection{Comparison issues} \label{section:comparisonissues}


\begin{table}
  \caption{When studying the effect of certain binary parameters, should we keep either the total cluster mass $\mcl$, or the number of particles $N=S+B$ constant? Below, we show the values of $\sigmalos$ and $\eta$ for three different models. Model~S1 is identical to model~S, except that no binaries are present. Models~S2 and S3 have $\binfrac=100\%$. Model~S2 has the same total mass as model~S1, while model~S3 has the same number of particles as model~S1. Each single/primary star has a mass $M=1\msun$. Each binary has $q=1$, $e=0$ and $a=10^3\rsun$. Columns 2--4 list the number of particles ($N=S+B$), the binary fraction, and the total mass of each cluster. Columns 5 and 6 list the measured velocity dispersion $\sigmalos$ of the stars (i.e., singles and binary components) and of the centre-of-mass of the particles $\sigmapart$. The corresponding values $\eta$ derived from these are listed in columns 7 and 8. As models~S1 and S3 have the same particle velocities $\sigmapart$, we choose to keep $\mcl$ (rather than $N$) fixed when comparing different clusters. Note that the values in column~8 are slightly larger than the canonical value of 9.75; see \S~\ref{section:aperture} for details.
    \label{table:comparison_issues}  }
  \begin{tabular}{lc cc cc cc}
    \hline \hline
    1 & 2 & 3 & 4 & 5 & 6 & 7 & 8 \\
    \hline 
    \#& $N$       & $\binfrac$ & $\mcl$   & $\sigmalos$ & $\sigmapart$ & $\eta$  & $\eta_{\rm part}$   \\          
     &            & \%         & $\msun$  & \kms  & \kms  & & \\
    \hline    
    S1&$10\,000$   & 0        & $10\,000$  & 0.92 & 0.92 & 10.23 & 10.23  \\
    S2&$10\,000$   & 100      & $20\,000$  & 7.10 & 1.29 &  0.34 & 10.24  \\
    S3&$ 5\,000$   & 100      & $10\,000$  & 7.06 & 0.91 &  0.17 & 10.30  \\
    \hline \hline
  \end{tabular}
\end{table}


In order to describe the effect of varying each binary parameter on the dynamical mass derivation, a comparison between models ``before'' and ``after'' the modification is necessary, whilst keeping all other parameters the same. This requirement, however, leads to some ambiguities. In particular, when adding binaries to a cluster, one has to make a decision whether to keep (i) the {\em cluster mass} $\mcl$ or (ii) the {\em number of particles} $N$ constant. By simply adding binary companions to a stellar population, the average mass of each particle increases, resulting in a larger total cluster mass. The cluster mass is then given by  $\mcl = N \langle M_1 \rangle (1+ \langle q\rangle \binfrac)$, where $\langle M_1 \rangle$ is the average primary/single mass, $\binfrac$ the binary fraction, $\mcl$ the total cluster mass and $\langle q\rangle$ the average mass ratio (assuming a mass ratio distribution that is independent of primary mass). Due to the larger mass, the particles move faster, such that $\sigmalos$ is larger. 


On the other hand, one could also decide to scale the number of particles $N$ in the cluster with binaries, such that its total mass $\mcl$ is equal to a cluster without binaries. For a binary population with a mass ratio distribution that is independent of primary mass, the number of particles  $N$ is given by $N=\mcl \langle M_1 \rangle^{-1} (1+ \langle q\rangle \binfrac)^{-1}$.


One thus has to make the choice to keep either $N$ or $\mcl$ constant. We use simulations to show the consequences of either choice. We perform our simulations with three models, S1--S3, for which the properties are listed in Table~\ref{table:comparison_issues}. Each model has a half-mass radius of 5~pc, and all stars have a mass of $1\msun$. Model~S1 is our reference model, which consists of $N=10\,000$ single, equal-mass stars, with a total mass of $10\,000\msun$. Models~S2 and~S3 include binaries. The binary fraction in these models is 100\%, and each binary has $a=10^3\rsun$, $e=0$, and $q=0$.  Model~S2 has the same number of particles $N$ as model~S1, while model~S3 has the same total mass $\mcl$ as model~S1. 


Model~S2 has a larger velocity dispersion than model~S1 for two reasons, (i) the presence of binaries, and (ii) the increased total mass. In order to separate the effects of these two changes, we calculate $\sigmalos$ and $\eta$ not only for the stars, but also for the (unmeasured) motion of the centres-of-mass of the binaries. Model~S3 has a similar mass, but a smaller number of particles.  The velocity of each particle, however, is very similar to that of model~S1.


Considering the motion of the centre-of-mass of each binary, it is better to study the effect of binarity using model~S3 rather than model~S2, i.e., to keep the total mass $\mcl$ constant in the comparison, rather than the number of particles $N$. Unless stated otherwise, we therefore keep the cluster mass $\mcl$ constant in each comparison, for the remainder of this paper.


\section{Dependence on cluster properties} \label{section:clusterproperties}

In this section we study the effect of varying the cluster properties on the dynamical mass determination. We discuss varying the virial ratio in \S\,\ref{section:virialratio}, the mass distribution and aperture size in \S\,\ref{section:aperture}, the number of particles in \S\,\ref{section:numberofparticles}, and the half-mass radius in \S\,\ref{section:halfmassradius}.


\subsection{The virial ratio $Q$} \label{section:virialratio}

Spitzer's equation assumes that a cluster is in virial equilibrium. Observations, however, suggest that many clusters form out of virial equilibrium \citep[e.g.,][]{bastiangoodwin2006}. This often results in early dissolution into the field star population (infant mortality) or significant mass loss (infant weight loss); see \cite{degrijs2007review} and references therein.

The virial ratio $Q \equiv -E_K/E_P$ of a star cluster is defined as the ratio between its kinetic energy $E_K$ and potential energy $E_P$. Clusters with $Q=0.5$ are in virial equilibrium, and those with $Q<0.5$ and $Q>0.5$ are contracting and expanding, respectively. Since $E_K \propto \sigmalos^2$, a star cluster has $\sigmalos^2 \propto Q$. A more generalized version of Spitzer's equation, including the effect of virial equilibrium, is therefore
\begin{equation}
  \mdyn = (2Q)^{-1}\,\eta \, \frac{\halflight  \sigmalos^2}{G} \,,
\end{equation}
which, if $Q=0.5$, reduces to Eq.~(\ref{equation:spitzer}). If a cluster is assumed to be in virial equilibrium while in reality it is expanding, the dynamical ass overestimates the true mass by a factor of $2Q$.

The most important reason that many (if not all) clusters are formed out of virial equilibrium, is that the star-forming efficiency is not 100\%. After removal of the gas by the winds of the most massive stars, the gravitational potential of the cluster is reduced significantly, which results in cluster expansion \citep[e.g.,][]{kroupaboily2002,bastiangoodwin2006}. \cite{goodwinbastian2006} define the {\em effective} star-forming efficiency (eSFE) $\epsilon$ as the star-forming efficiency that one would derive from the virial ratio under the assumption that the star-forming cloud was originally in virial equilibrium: $Q = (2\epsilon)^{-1}$. Under this assumption, the dynamical mass overestimation for a cluster of single stars is $\mdyn/\mcl=\epsilon^{-1}$. 


\subsection{The mass distribution and aperture size} \label{section:aperture}


\begin{figure}[!bt]
  \centering
  \includegraphics[width=0.48\textwidth,height=!]{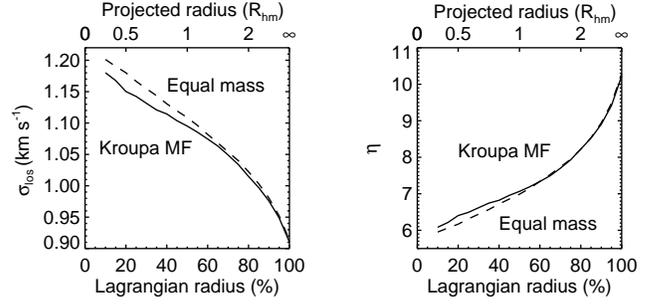}
  \caption{The effect of the aperture size on $\sigmalos$ and $\eta$, for model~S (dashed curves) and model~R (solid curves). Each model has a zero binary fraction, a half-mass radius $\halfmass=5$~pc, and a total mass $\mcl=10^4\msun$. This figure shows the derived $\sigmalos$ and $\eta$ for all stars {\em within} an aperture, as a function of the aperture size. The derived value $\eta$ for each model depends only mildly on the mass distribution, while the aperture size is of much greater importance. Depending on the size of the aperture, the total mass of a star cluster may be overestimated by a factor of two, if this selection effect is not taken into account. 
    \label{figure:effect_aperture} }
\end{figure}


Spitzer derived Eq.~(\ref{equation:spitzer}) for a cluster of equal-mass single stars, assuming that the line-of-sight velocity is measured precisely at the half-mass radius. This measurement is impractical to do; usually an integration is performed over a large part of the cluster. In reality the outskirts of a cluster are dominated by the background field star population, so that the analysis is restricted to the inner part of the cluster. These selective measurements introduce biases. The stellar density in the cluster centre is higher than that in the outskirts, resulting in a larger velocity dispersion. The magnitude of this difference (and thus the difference in $\eta$) depends on the density profile of the cluster. If this selection effect is not taken into account, the derived $\mdyn$ depends on which part of the cluster is observed. 


Below, we study the dynamical mass overestimation as a function of position in two ways. First, we study the overestimation if the velocity dispersion is measured at a projected radius $\rho$ from the cluster centre, as a function of $\rho$. Secondly, we study the dynamical mass overestimation for different apertures of integration, as a function of the aperture size. The former can be calculated analytically for Plummer models, while the latter can be directly compared with observations.


The line-of-sight velocity dispersion of the centres-of-mass at a certain projected distance $\rho$ from the cluster centre is
\begin{equation} \label{equation:lagrangiansigma}
  \sigmapart^2(\rho) = \frac{3\pi}{64}\frac{G\mcl}{\halfmass} \left( 1 + \frac{\rho^2}{\halfmass^2} \right)^{-1/2}
\end{equation}
for a Plummer model \citep[e.g.,][]{heggiehut}. At the half-mass radius, $\rho=\halfmass$, this equation simplifies to 
\begin{equation}
  \sigmapart(\halfmass)
  = 
  2.116 \ 
  \left( \frac{\mcl}{10^4 \msun} \right)^{\tfrac{1}{2}}
  \left( \frac{\halfmass}{\mbox{pc}} \right)^{-\tfrac{1}{2}} 
  \ \ \mbox{km\,s}^{-1}  \,.
\end{equation}
Note that this is the projected velocity dispersion at projected radius $\rho$, rather than the integrated value within a radius $\rho$. The dynamical mass may be significantly overestimated if this selection effect is not taken into account. An expression for the dynamical mass overestimation as a function of $\rho$ is obtained by substituting Eq.~(\ref{equation:lagrangiansigma}) into Eq.~(\ref{equation:spitzer}):
\begin{equation}
  \frac{\mdyn}{\mcl} \approx \sqrt{2} \ \left( 1 + \frac{\rho^2}{\halfmass^2} \right)^{-1/2} \,.
\end{equation}
For velocity dispersions measured in the cluster centre this results in a mass overestimation by $\sim 40\%$. Measurements at the half-mass radius provide the correct $\mdyn$, while measurements in the cluster outskirts result in an underestimation of the mass. 


In reality, the velocity dispersion of a star cluster is determined from the measurements within an aperture. The size of the aperture is usually defined by the projected radius at which the projected stellar density becomes so low that the contribution of background stars dominates the brightness and velocity dispersion. 


We illustrate the dependence of the derived dynamical mass on the aperture size by performing simulations of models~S and~R, each with a zero binary fraction, $\binfrac=0\%$. We use $N=10\,000$ for model~S and $N=28\,000$ for model~R, so that each cluster has a mass of $\mcl=10^4\msun$. Fig.~\ref{figure:effect_aperture} shows $\sigmalos$ and $\eta$ as a function of the Lagrangian radius $\rlag$ within which the measurement is performed. The (projected) $n$th-percentile Lagrangian radius $\rlag$ is defined as the radius which includes a fraction $f$ of the total mass of a cluster. For example, the projected radius which contains 50\% of the mass, $R_{50\%}$ (i.e., the half-mass radius $\halfmass$), for example, is found by solving $M(\rho)/\mcl=50\%$. For a Plummer model, the mass $M(\rho)$ within a projected radius $\rho$ is given by 
\begin{equation}
  M(\rho) = \mcl \, \left( 1 + \frac{\halfmass^2}{\rho^2} \right)^{-1}
\end{equation}
\citep[e.g.,][]{heggiehut}. The projected radius $\rlag$ which contains a fraction $f$ of the cluster mass, is given by
\begin{equation}
  \rlag = \frac{ \halfmass }{ \sqrt{ f^{-1} -1 } } \,.
\end{equation}


The choice of the aperture size may result in dynamical mass measurements differing by up to a factor of two. For example, if only the stars within a radius $R_{20\%} = \tfrac{1}{2}\halfmass$ are observed, while this is not taken into account, the total mass is overestimated by $\sim 60\%$, for each model. The velocity dispersion of the model with the Kroupa mass distribution is slightly lower than that of the equal-mass model in the cluster centre, while the values are virtually the same for the cluster as a whole. The derived values of $\eta$ for the Kroupa model are 3\% larger than those of the equal-mass model in the cluster centre, but practically equal if the entire cluster is taken into account. This indicates that Spitzer's equation is not very sensitive to the mass distribution.


\subsection{The number of particles $N=S+B$} \label{section:numberofparticles}


Under the assumptions made by Spitzer, Eq.~(\ref{equation:spitzer}) is independent of the number of particles. The only relation between $\mdyn$ and $N$ has a statistical nature: if the number of particles is small, the statistical error on $\mdyn$ is large.
If binaries are present in the star cluster, however, the derived dynamical mass decreases with increasing $N$, up to the point when $\sigmalos$ is dominated by the particle motion. This can be understood as follows. Imagine a group of binary systems, which have a given orbital velocity. This orbital velocity is independent of the number of stars in a cluster. The stellar density (which is proportional to $N$), however, affects the motion of the centre-of-mass of each binary system. The particles move faster in a dense cluster, while the orbital motion of each star in a binary system remains the same. The contribution of $\sigmapart$ to $\sigmalos$, relative to $\sigmaorb$, thus becomes smaller; the dynamical mass overestimation is less severe for clusters with larger $N$.


An open star cluster typically has a mass of $10^4\msun$. Assuming a Kroupa mass distribution and a minimum mass of $0.02\msun$, the number of particles (singles or binaries) in such a cluster is expected to be $N \approx (1.4-2.8) \times 10^4$. The upper limit represents a cluster without binary systems, and the lower limit a cluster with $\binfrac=100\%$ and a mass ratio of unity for all binaries. For young massive star clusters, with a total mass of order $10^6\msun$, the number of particles is expected to be $N=(1.4-2.8)\times10^6$. 

\begin{figure}[!bt]
  \centering
  \includegraphics[width=0.48\textwidth,height=!]{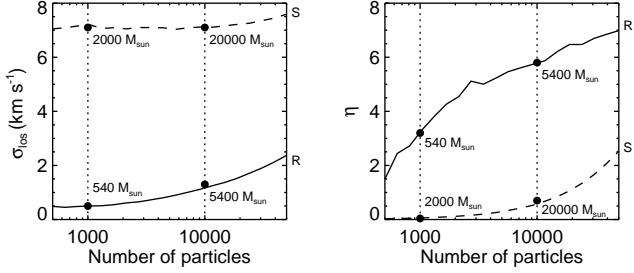}
  \caption{The effect of the number of particles $N=S+B$ (horizontal axis) in the cluster, for model~S (dashed curves) and model~R (solid curves). Each model has a binary fraction of 100\% and a half-mass radius $\halfmass=5$~pc. For a given number of particles $N$, model~S is $3.69$~times more massive than model~R. In dense clusters (with large $N$) the motion of the binaries in the cluster potential dominates $\sigmalos$, so that $\eta$ is close to its ``zero binary fraction'' value of 9.75. In sparse clusters (with low $N$), $\sigmalos$ is dominated by the orbital motion of the binaries. In this case, the overestimation of $\mdyn$ is largest.  
\label{figure:effect_SR_N} }
\end{figure}

In order to study the relationship between $N$ and the derived $\eta$ and $\mdyn$, we simulate models~S and~R, and evaluate the results for different values of $N$. Each model has a binary fraction of 100\% and a half-mass radius $\halfmass=5$~pc. 
%
%
The results are shown in Fig.~\ref{figure:effect_SR_N}. Due to the different mass and mass ratio distributions of each model, the average mass of a particle is different: $2\msun$ in model~S and $0.54\msun$ in model~R. Given the number of particles $N$, model~S is therefore $3.69$ times more massive than model~R. This difference in total mass contributes to the larger velocity dispersion in model~S, which is reflected in Fig.~\ref{figure:effect_SR_N}.


According to Eq.~(\ref{equation:spitzer}), $\eta$ is proportional to $\mdyn\sigmalos^{-2}$ for a set of clusters with given half-mass radius $\halfmass$. Since $\mdyn\propto N$, this can be rewritten as $\eta \propto N\sigmalos^{-2}$.  Eq.~(\ref{equation:spitzer}) applies for a particle-dominated cluster, irrespective of $N$. In a binary-dominated cluster, the measured velocity dispersion  $\sigmalos$ is independent of $N$. For binary-dominated clusters we thus have  $\eta \propto N$. Ignoring the presence of binaries in a binary-dominated cluster therefore results in a mass overestimation $\mdyn/\mcl \propto N^{-1}$. Note that this relation is only applicable in the binary-dominated regime, and therefore cannot be extended to arbitrarily large $N$.


These properties are clearly shown in Fig.~\ref{figure:effect_SR_N}. Note that, since we adopt a fixed half-mass radius of $5$~pc, model~S is only binary-dominated for $N \la 10\,000$. In this regime, model~S shows the predicted constant $\sigmalos$, and the corresponding $\eta \propto N$.
Model~R, on the other hand, shows a stronger correlation between $\sigmalos$ and $N$; the overestimation of $\mdyn$ is also less severe.


We finally describe the subtle difference between varying the number of particles $N$ and the cluster mass $\mcl$. These quantities are related as $\mcl = N \langle M_T \rangle$, where $\langle M_T \rangle$ is the average particle (single or binary) mass. When changing $\mcl$ while keeping $N$ fixed, one changes the average mass $\langle M_T \rangle$ of the binaries. The latter is equivalent to changing the mass distribution, the mass ratio distribution, or the binary fraction. We refer to \S\,\ref{section:aperture}, \S\,\ref{section:massratio} and \S\,\ref{section:binaryfraction} for a discussion on these issues, respectively.

\subsection{The half-mass radius $\halfmass$} \label{section:halfmassradius}


\begin{figure}[!bt]
  \centering
  \includegraphics[width=0.48\textwidth,height=!]{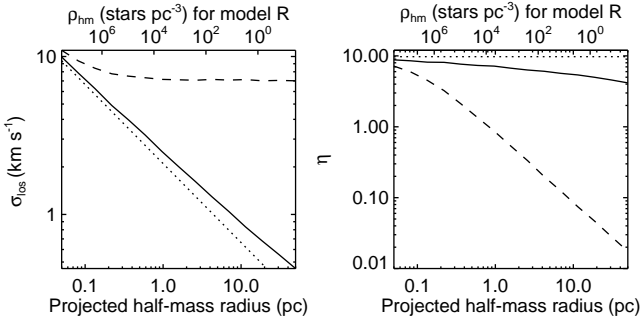}
  \caption{The parameters $\sigmalos$ and $\eta$ as a function of the half-mass radius $\halfmass$, for model~S (dashed curves) and model~R (solid curves). The dotted curve indicates the results for a cluster without binaries. Each model has $\binfrac=100\%$ and $\mcl=10^4\msun$. Above both panels we indicate the average stellar density within the half-mass radius. As model~S is binary-dominated, we find, to first order approximation, $\sigmalos \approx$~constant and $\eta \propto \halfmass^{-1}$. The effect of binaries is less pronounced for model~R.
\label{figure:effect_SR_halfmass} }
\end{figure}


Given the number of particles $N$ or the cluster mass $\mcl$, the velocity at which the particles move is dominated by the size $\halfmass$ of the star cluster. Particles move fast in small (dense) clusters, and slowly in large (sparse) clusters. The orbital motion of binary components, however, is unaffected by the size of the cluster. In a large (sparse) cluster, we therefore expect the binary orbital motion to dominate $\sigmalos$, and in a small cluster the motion of the particles dominates. The dynamical mass overestimation from Spitzer's equation is thus least severe for small (dense) clusters.


We study the relation between $\halfmass$, $\sigmalos$ and $\eta$ by simulating clusters of different size. We adopt the properties of models~S and~R, each with a binary fraction of 100\% and a cluster mass of $10^4\msun$, and study clusters with sizes in the range $0.01~\mbox{pc} <\halfmass<100~\mbox{pc}$. The resulting trend of $\sigmalos$ and $\eta$ with $\halfmass$ is shown in Fig.~\ref{figure:effect_SR_halfmass}. Above each panel we indicate, $\langle \rho \rangle_{\rm hm}$, the average stellar density within the half-mass radius for model~R, given by Eq.~(\ref{equation:density}).


The figure shows that, as expected, larger clusters have smaller values for $\sigmalos$ and $\eta$. The left-hand panel shows that $\sigmalos$ for model~S (dashed curve) is approximately constant for $\halfmass \ga 0.5$~pc, indicating that such clusters are binary-dominated. For these clusters the orbital motion is independent of $\halfmass$. In Spitzer's equation, $\eta \propto \halfmass^{-1}$, which is indeed observed in the right-hand panel of Fig.~\ref{figure:effect_SR_halfmass}. For clusters with $\halfmass \la 0.5$~pc, on the other hand, $\sigmalos$ decreases with increasing $\halfmass$, indicating that the centre-of-mass motion of the particles is not negligible as compared to the orbital motion; these clusters are of the intermediate case. 


The effect of binaries is less pronounced in model~R (solid curves in Fig.~\ref{figure:effect_SR_halfmass}). The value of $\sigmalos$ is affected by both the particle motion and the orbital motion, but mostly dominated by the former. If $\sigmalos$ were completely dominated by the particle motion one would expect $\sigmalos \propto \halfmass^{-1/2}$ and $\eta =$~constant. In Fig.~\ref{figure:effect_SR_halfmass} this is, to first order, the case, although the dynamical mass of clusters with large $\halfmass$ is clearly overestimated, if the assumption of a zero binary fraction is made. For clusters with $\mcl=10^4\msun$ and $\halfmass=10$~pc, for example, we find $\eta \approx 5$. This is half the canonical value of $\eta$. Ignoring binaries in such a cluster may thus result in a dynamical mass overestimation by a factor of two.


\section{Dependence on the properties of the binary population} \label{section:binarypopulationproperties}

In this section we discuss the influence of varying the different binary parameters on the dynamical mass determination. We discuss variations of the semi-major axis (and period) distribution in \S\,\ref{section:semimajoraxis}, the mass ratio distribution in \S\,\ref{section:massratio}, the eccentricity distribution in \S\,\ref{section:eccentricity}, and the binary fraction in \S\,\ref{section:binaryfraction}.


\subsection{The semi-major axis distribution $f_a(a)$} \label{section:semimajoraxis}


\begin{figure}[!bt]
  \centering
  \includegraphics[width=0.48\textwidth,height=!]{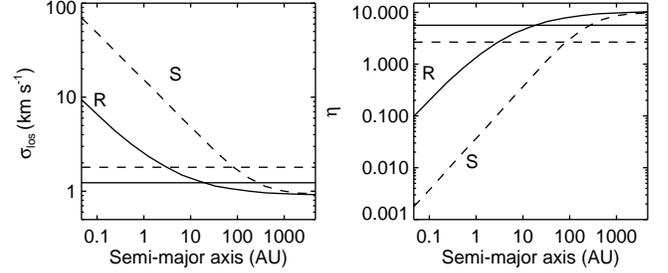}
  \caption{The parameters $\sigmalos$ and $\eta$ for models with different semi-major axis distributions $f_a(a)$, for model~S (dashed curves) and model~R (solid curves). Each model has a binary fraction $\binfrac=100\%$, a half-mass radius $\halfmass=5$~pc and a total mass $\mcl=10^4\msun$. The curves indicate the results for models where {\em all} binaries have identical semi-major axis $f_{a_0}(a)=\delta(a-a_0)$, with $a_0$ indicated on the horizontal axis. In the binary-dominated case ($a_0 < 10$~au), we have, to first order approximation, $\eta \propto a_0$. The horizontal lines indicate the results for a distribution $f_{\rm Opik}(a) \propto a^{-1}$ (\opik's law). The results for the log-normal period distribution $f_{\rm DM}(P)$ (not shown) are practically indistinguishable from those for \opik's law. 
    \label{figure:effect_SR_a} }
\end{figure}


The distribution of binary semi-major axes (or orbital periods), relative to the size of the cluster is one of the most important parameters affecting our interpretation of the observed velocity dispersion $\sigmalos$. The orbital velocity of a binary system is proportional to $a^{-1/2}$. 

In order to only extract the contribution of $a$ to the mass derivation, we simulate clusters with varying semi-major axis distributions. We first study the results for clusters in which all binaries have an {\em identical} semi-major axis: $f_{a_0}(a)=\delta(a-a_0)$. Each model has a binary fraction of 100\%, a half-mass radius of 5~pc, and a total mass of $10^4\msun$. The dependence of $\sigmalos$ and $\eta$ on the value of $a_0$ is shown in Fig.~\ref{figure:effect_SR_a}, for both models~S and~R. In a binary-dominated cluster $\sigmalos \approx \sigmaorb$, we therefore have $\sigmalos \propto a^{-1/2}$, and hence $\eta \propto a^{-1}$. This is clearly shown for model~S (dashed curves) in  Fig.~\ref{figure:effect_SR_a}. The effect is less pronounced for model~R, which is neither binary-dominated nor particle-dominated.


In reality, stellar groupings contain binaries with a large range of orbital sizes, unlike the example described above. These orbital sizes can be quantified using a semi-major axis distribution $f_a(a)$, or, indirectly, an orbital period distribution $f_P(P)$. 
The flat distribution in $\log a$, commonly known as \"{O}pik's law, has been observed for a wide range of stellar populations \citep[e.g.,][]{opik1924,vanalbada1968,vereshchagin1988,poveda2004,poveda2007}, and is equivalent to
\begin{equation} \label{equation:opikslaw}
f_a(a) \propto a^{-1} 
\quad \quad \amin \leq a \leq \amax \,,
\end{equation}
with $\amin \approx 10\rsun$ and $\amax \approx 0.02$~pc ($4500$~au). \cite{duquennoy1991} studied binarity among solar-type stars in the solar neighbourhood  and found a log-normal period distribution: 
\begin{equation} \label{equation:duquennoyperiods}
f_{\rm DM}(P) \propto \exp \left\{ - \frac{(\log P -  \overline{\log P} )^2 }{ 2 \sigma_{\log P}^2 }  \right\} 
\quad \quad \pmin \leq P \leq \pmax \,,
\end{equation}
with $\pmin \approx 4$~days and $\pmax \approx 0.3$~Myr. They find $\overline{\log P} = 4.8$, $\sigma_{\log P} = 2.3$, where $P$ is in days. The latter distribution is often used as the standard reference for the orbital size distribution of a binary population. The size of the smallest orbits is determined by the radii of the stars or, more precisely, by the semi-major axis at which Roche lobe overflow occurs.  The size of the largest orbit is determined by properties of the surrounding stellar population, in particular the stellar density. In our models with the semi-major axis distribution $f_a(a)$ we adopt the limits $\amin = 10\rsun$ ($\approx 0.05$~au) and $\amax=10^6\rsun$ ($\approx 0.02$~pc). In the models with the period distribution $f_{\rm DM}(P)$, we adopt $\pmin=4$~days and $\pmax=0.3$~Myr. The results for \opik's law and the log-normal period distribution are indicated by the horizontal lines in Fig.~\ref{figure:effect_SR_a}. Both distributions show practically indistinguishable results.

The upper limits $\amax$ and $\pmax$ are in reality dependent on the environment, in particular on the stellar density \citep[e.g.,][]{bahcall1985,close1990,chaname2004}. Given the half-mass radius $\halfmass$ and number of particles $N$ in a cluster, there is a maximum semi-major axis at which a binary system is marginally stable. Very wide binary systems are quickly ionised, as their binding energy is too weak to keep the components together. Below, we derive a simple expression for the maximum semi-major axis $\amax$, relating it to $N$ and $\halfmass$.

Consider a star cluster with $N$ particles and a half-mass radius $\halfmass$. The $\tfrac{1}{2}N$ particles within the half-mass radius occupy a total volume $\tfrac{4}{3}\pi\halfmass^3$. The average volume available to one particle equals $\tfrac{2}{3}\pi\halfmass^3 N^{-1}$, which is equivalent to a sphere of radius $(2N)^{-1/3}\,\halfmass$. The maximum semi-major axis is thus expected to be of order $\amax \approx (2N)^{-1/3}\,\halfmass$. Note that this is a conservative estimate as the density in the cluster centre is significantly higher than at the half-mass radius; the maximum semi-major axis is likely smaller than $\amax$. The semi-major axis distribution is further truncated for older clusters due to dynamical evolution. 

Our simulations show that the effect of our $\amax=0.02$~pc assumption is relatively small. We have compared the difference between clusters with $\amax=0.02$~pc, and those with the more realistic $\amax \approx (2N)^{-1/3}\,\halfmass$ for a cluster with half-mass radius $\halfmass$ and $N$ particles. The difference in the mass-overestimation is $\Delta (\mdyn/\mcl) \approx 4-6\%$ for realistic values of $\amax$.


\subsection{The mass ratio distribution $f_q(q)$} \label{section:massratio}


\begin{figure}[!bt]
  \centering
  \begin{tabular}{c}
    \includegraphics[width=0.48\textwidth,height=!]{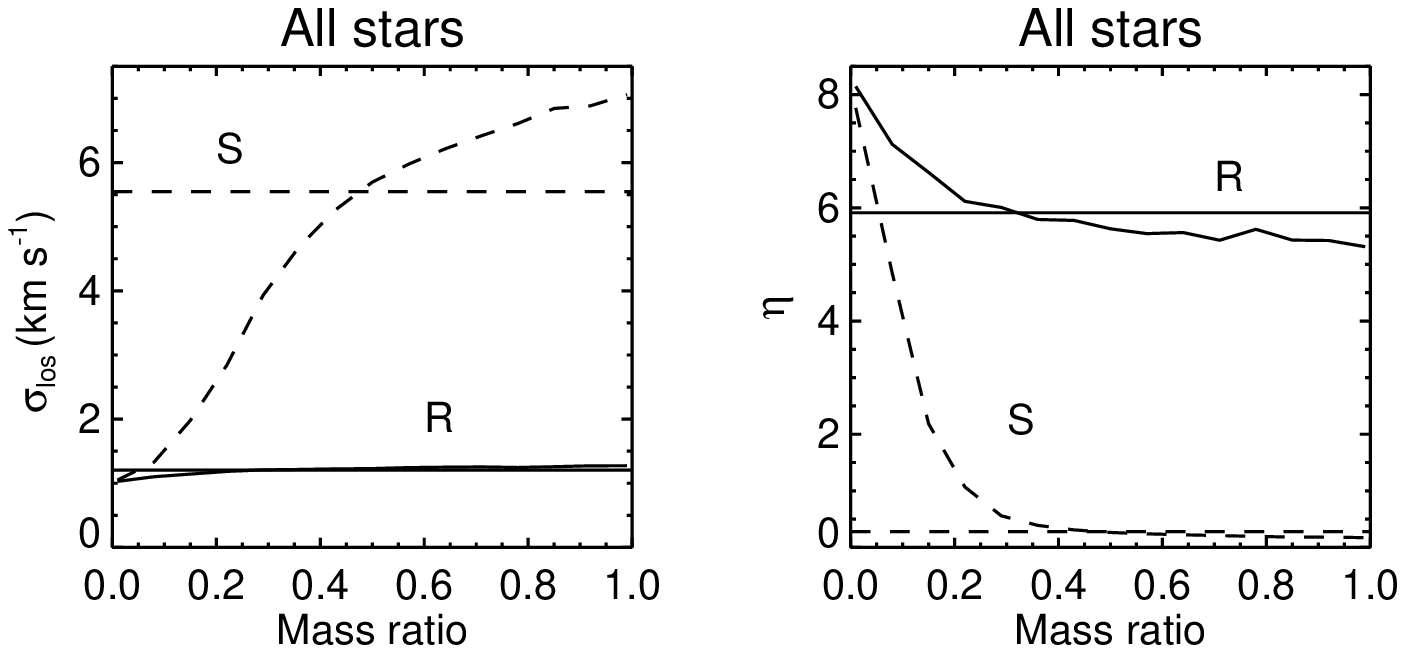}\\
    \includegraphics[width=0.48\textwidth,height=!]{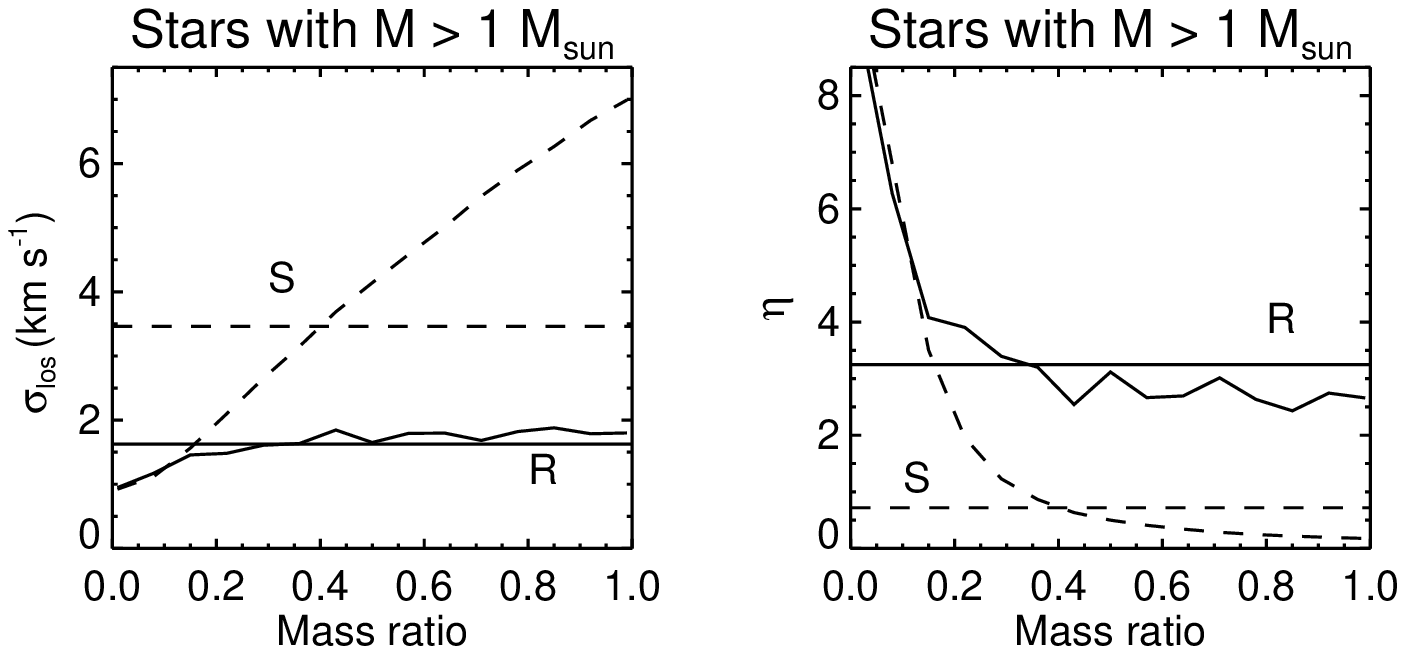}
  \end{tabular}
  \caption{The values $\sigmalos$ and $\eta$ for model~S (dashed curves) and~R (solid curves), as a function of the mass ratio distribution $f_q(q)$. Each model has $\binfrac=100\%$, $\halfmass=5$~pc, and $\mcl=10^4\msun$. The curves indicate the results for models in which all binaries have a fixed mass ratio $q_0$, which is indicated along the horizontal axis. The top panels show the results for {\em all} stars in the cluster. The bottom panels show the results for the stars more massive than $1\msun$ (i.e., mostly primary stars). The horizontal lines in each panel show the results for the flat mass ratio distribution $f_{q}(q) =1$. For model~R with a flat mass ratio distribution, the dynamical mass overestimation is $\mdyn/\mcl \approx 1.7$ if all stars are included in the fit, and $\mdyn/\mcl\approx 3$ if only the stars with $M > 1\msun$ are included.
\label{figure:effect_SR_q} }
\end{figure}


In this section we study the relation between the mass ratio distribution $f_q(q)$ and the systematic error in $\mdyn$ caused by neglecting the binaries.
For each model we first adopt a mass ratio distribution $f_{q_0}(q) = \delta(q-q_0)$, and evaluate $\eta$ for $0<q<1$; see the top panels in Fig.~\ref{figure:effect_SR_q}. The figure indicates that $\sigmalos$ increases with increasing $q_0$ for model~S. This effect is less prominent for model~R. The value of $\eta$ decreases with increasing $q_0$ for both models, which is most pronounced for small mass ratios. For very small mass ratios $\eta$ reaches 9.75, the value for a cluster without binaries. The mass overestimation for model~R is approximately constant for $q_0 \ga 0.3$.


In reality, a cluster contains binary systems covering the full range $0<q\leq 1$. We therefore also perform our simulations using a continuous mass ratio distribution $f_q(q)$. The mass ratio distribution has been studied for various stellar populations. For example, \cite{kouwenhoven_recovery} find a distribution $f_q(q)\propto q^{-0.4}$ for intermediate-mass stars in the nearby OB~association Sco~OB2. \cite{kobulnicky2007} find very similar results for Cyg~OB2. Others find a large number of $q\approx 1$ (``twin'') binaries, mostly among massive stars \citep[e.g.,][]{pinsonneault2006,lucy2006,soderhjelm2007}. In this paper we study the results for a flat mass ratio distribution $f_{\rm flat}(q) = 1$. This distribution is solely chosen as an example, in order to show the (typical) effect of having a mass ratio distribution rather than a single mass ratio. The results for this mass ratio distribution are indicated by the horizontal lines in Fig.~\ref{figure:effect_SR_q}. For both models~S and~R, the mass overestimation is similar to that of a model with $q_0 \approx 0.4$; the distribution $f_q(q)=1$ can thus be described with an effective mass ratio $\qeff \approx 0.4$.


In an unequal-mass binary system the most massive star orbits with the smallest velocity. In general, the velocity dispersion measured from spectral lines reflects the properties of a specific subset of the cluster members only. These are often bright stars with narrow lines. As the radial velocity amplitude of a primary star increases with increasing mass ratio, the value of $\sigmalos$ increases as well. The bottom panels in Fig.~\ref{figure:effect_SR_q} illustrate this effect. In this example only the stars more massive than $1\msun$ (i.e., mostly primary stars) are included. We have selected a subset of more massive stars. The orbital velocity of binaries follows $v_{\rm orb} \propto M_T^{-1/2}$. On average, the contribution of $\sigmaorb$ to $\sigmalos$ is therefore larger. We are thus measuring a smaller $\eta$ and derive a larger dynamical mass overestimation by looking at the more massive stars. Not taking this selection effect into account may result in a further overestimation of the dynamical mass of the star cluster. In this example we find that by selecting the stars with $M>1 \msun$, the mass is overestimated by a factor of $\sim 1.8$ more than if we would have selected all stars: $(\mdyn/\mcl)_{\rm M>1 \msun} \approx 1.8\,(\mdyn/\mcl)_{\mbox{\scriptsize all stars}}$. 


\subsection{The eccentricity distribution $f_e(e)$} \label{section:eccentricity}


\begin{figure}[!bt]
  \centering
  \begin{tabular}{c}
    \includegraphics[width=0.48\textwidth,height=!]{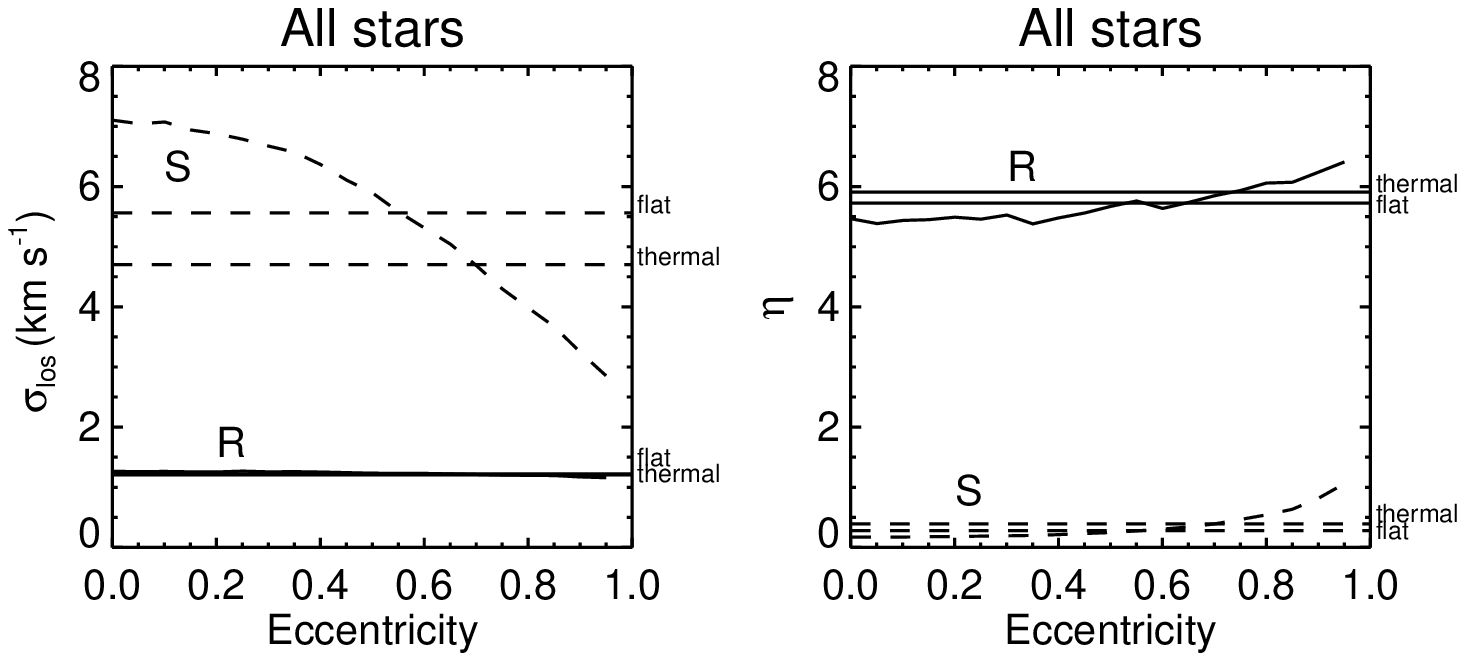}\\
    \includegraphics[width=0.48\textwidth,height=!]{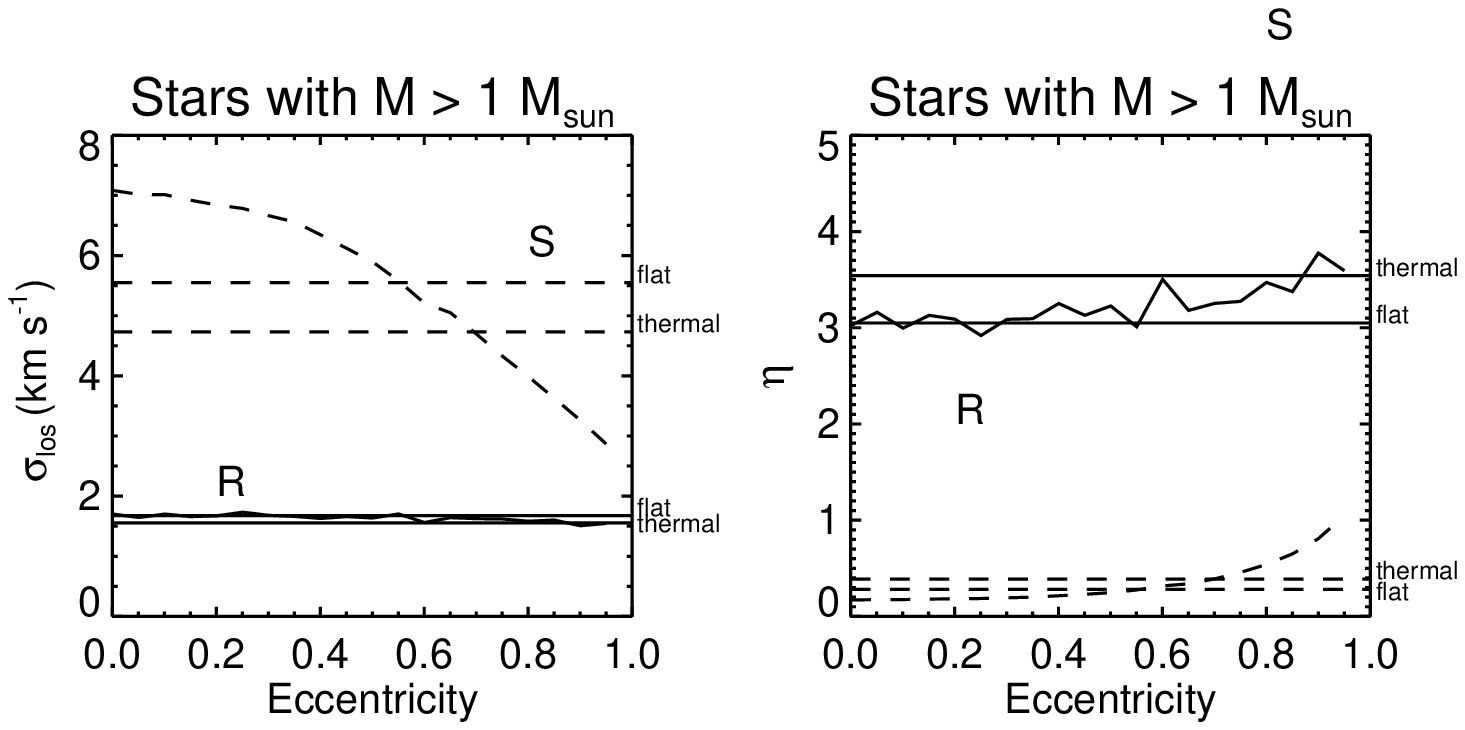}\\
  \end{tabular}
  \caption{The results for different eccentricity distributions, for model~S (dashed curves) and model~R (solid curves). Each model has $\mcl=10^4\msun$, $\halfmass=5$~pc and $\binfrac=100\%$. The curves indicate the results for models with an eccentricity distribution $f_{e_0}(e)=\delta(e-e_0)$, where $e_0$ is indicated along the horizontal axis. The horizontal lines indicate the results for the flat distribution $f_{\rm flat}(e)=1$ and the thermal distribution $f_{2e}(e)=2e$.  {\em Top}: results for all stars. Binaries with a large eccentricity spend most of their time near apastron, where their velocity is relatively low. The dynamical mass derivation for clusters with a large average eccentricity is therefore less affected by the presence of binaries. For realistic models, however, the effect of varying the eccentricity on $\eta$ is small as compared to changes in the other binary parameters, such as the binary fraction and the semi-major axis distribution. {\em Bottom}: results for stars with masses greater than $1\msun$ (i.e., mostly the primaries). With additional selection effect, the dynamical mass overestimation increases significantly: $(\mdyn/\mcl)_{\rm M>1 \msun} \approx 1.8\,(\mdyn/\mcl)_{\mbox{\scriptsize all stars}}$.
\label{figure:effect_SR_e} }
\end{figure}


A binary system with eccentricity $e>0$ spends most of its time near apastron, where the orbital velocity of both components is relatively small. The probability of the binary near periastron (i.e., of finding a large velocity) is small. For this reason, the derived value for the dynamical mass decreases with increasing average orbital eccentricity.


In order to study the effect of varying the eccentricity on the derived dynamical mass, we perform simulations of models~S and~R. We evaluate the results for three eccentricity distributions: the single-value distribution $f_{e_0}(e)=\delta(e-e_0)$, the thermal distribution $f_{2e}=2e$, and the flat distribution $f_{\rm flat}(e)=1$. The thermal eccentricity distribution is expected from energy equipartition \citep{heggie1975}. This distribution is frequently adopted in dynamical models, for reasons of simplicity. \cite{duquennoy1991} show that observations of solar-type stars in the solar neighbourhood are consistent with $f_{2e}(e)$. For populations with a thermal eccentricity distribution, most binaries are in highly eccentric orbits (50\% have $e>0.7$). Binarity among these objects is particularly difficult to detect spectroscopically, as these binaries spend most of their time near apastron. Although most observations are consistent with the thermal distribution, they are often equally consistent with the flat eccentricity distribution $f_{\rm flat}(e)=1$, which is the third distribution considered in this paper.


The top panels in Fig.~\ref{figure:effect_SR_e} show the results for models~S and~R. The curves indicate the results for $f_{e_0}(e)$, where the eccentricity $e_0$ is indicated along the horizontal axis. The increasing $\eta$ for increasing average eccentricity is clearly visible. For any realistic star cluster (similar to model~R), however, the contribution to the systematic error in $\eta$ due to variations in $f_e(e)$ is small as compared to that of the other binary parameters, such as $f_a(a)$ and $\binfrac$. For these clusters, those with $e\equiv 0$ are associated with a mass overestimation of $\sim80\%$, while those with  $e\equiv 0.95$ are associated with a mass overestimation of $\sim50\%$. The horizontal lines indicate the results for $f_{2e}(e)$ and $f_{\rm flat}(e)$. For realistic models (e.g., model~R), the results for $f_{2e}(e)$ and $f_{\rm flat}(e)$ are very similar to those of the clusters with $f_{e_0}(e)$: the corresponding mass overestimation is $\mdyn/\mcl \approx 65-70\%$.


The bottom panels of Fig.~\ref{figure:effect_SR_e} show results for the same simulations, but now only for the subset of stars which are more massive than $1\msun$ (cf. Fig.~\ref{figure:effect_SR_q} and \S\,\ref{section:massratio}). The results for the distributions $f_{2e}(e)$ and $f_{\rm flat}(e)$ are again very similar to those of clusters with $f_{e_0}(e)$. For each model, the mass overestimation is now $~1.8$ times larger: $(\mdyn/\mcl)_{\rm M>1 \msun} \approx 1.8\,(\mdyn/\mcl)_{\mbox{\scriptsize all stars}}$; cf. \S\,\ref{section:massratio}. By selecting a subset of massive stars, the derived mass is thus significantly further overestimated. The reason that the mass selection effect results in a larger dynamical mass overestimation is that for the orbital velocities of binary stars $v_{\rm orb} \propto M_T^{-1/2}$, so that the measured $\sigmalos$ is larger. 


\subsection{The binary fraction $\binfrac$} \label{section:binaryfraction}


\begin{figure}[!bt]
  \centering
  \includegraphics[width=0.48\textwidth,height=!]{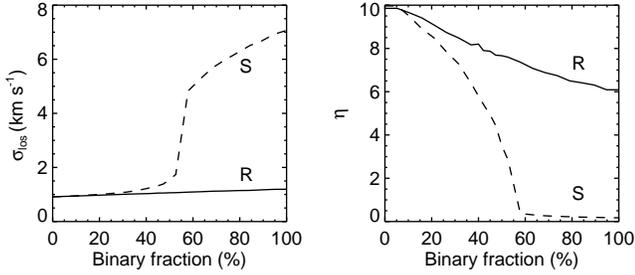}
  \caption{The effect of the binary fraction $\binfrac$ in the cluster, for model~S (dashed curves) and model~R (solid curves). The figure shows that, as expected, the overestimation of $\mdyn$ increases with increasing binary fraction. Model~S shows a strong increase in $\sigmalos$ at $\binfrac \approx 50\%$, while the effect for model~R is approximately linear.
\label{figure:effect_SR_fm} }
\end{figure}


In this section we discuss the relation between the dynamical mass overestimation $\mdyn/\mcl$ and the binary fraction $\binfrac$. Observations and simulations have indicated that the vast majority of stars, possibly even all stars, are formed in binary or multiple systems \citep[e.g.,][and references therein]{mathieu1994,mason1998,goodwinkroupa2005,kobulnicky2007,kouwenhoven_adonis,kouwenhoven_recovery}. The results presented in the sections above are based on clusters with a binary fraction of 100\%. These results are therefore reasonably accurate for young star clusters. The binary population in clusters is known to decrease with time: \cite{sollima2007} find $\binfrac = 10-50\%$ for 13~low-density globular clusters. The binary fraction in dense cores is even lower: \cite{cool2002}, for example, find $\binfrac < 7\%$ for the globular cluster NGC\,6397.


Clearly, if the binary fraction is negligible ($\binfrac \approx 0\%$), the star cluster can be considered as a cluster of single stars. In this case the value of $\eta$ for a single-star cluster can be used, and Eq.~(\ref{equation:spitzer}) gives a good approximation to the true mass. The presence of binaries gradually becomes more important if the binary fraction increases. When comparing two models with a different binary fraction, one has to make the choice between either keeping the total number of particles $N=S+B$ constant, or keeping the total mass $\mdyn$ of the cluster constant. For reasons described in \S\,\ref{section:comparisonissues} we choose to do the latter. 
 

The results of varying the binary fraction are shown in Fig.~\ref{figure:effect_SR_fm}, for model~S (dashed curves) and model~R (solid curves). For both models, the presence of binaries is relatively unimportant if the binary fraction is below $\sim 20\%$. For larger binary fractions, binaries rapidly become more important. 
For model~S, going from $\binfrac=40\%$ to $60\%$, the binaries suddenly become important for $\sigmalos$. The reason for this is that, for model~S, the velocity dispersion of the binary orbital motions is much larger than that of the particle motion. For a binary fraction $0\%<\binfrac<100\%$, the measured velocity distribution has two components: a broad component for the binary systems and a small component for the single stars. When fitting a Gaussian distribution to the (two-component) velocity distribution, the single stars dominate the fit for small $\binfrac$, while the binary systems dominate for large $\binfrac$. For non-linear least-square fitting of a Gaussian function, this results in a rapid increase of the best-fitting $\sigmalos$ when the binary fraction increases from 40\% to 60\%.
For model~R, on the other hand, the transition is much smoother, as the particle and orbital motions are comparable, and results in an approximately linear relation between $\binfrac$ and $\eta$. 
For model~R in our example, the relation between the mass overestimation and the binary fraction can be approximated by $\mdyn/\mcl \approx (1 - 0.36 \binfrac)^{-1}$. This is, however, not a general result; models with other properties should be studied individually, as $\mdyn/\mcl$ increases with decreasing $\mcl$, decreasing $N$ and increasing $\halfmass$.


\section{When can binaries be ignored?} \label{section:whencanbinariesbeignored}


\begin{figure}[!bt]
  \includegraphics[width=0.48\textwidth,height=!]{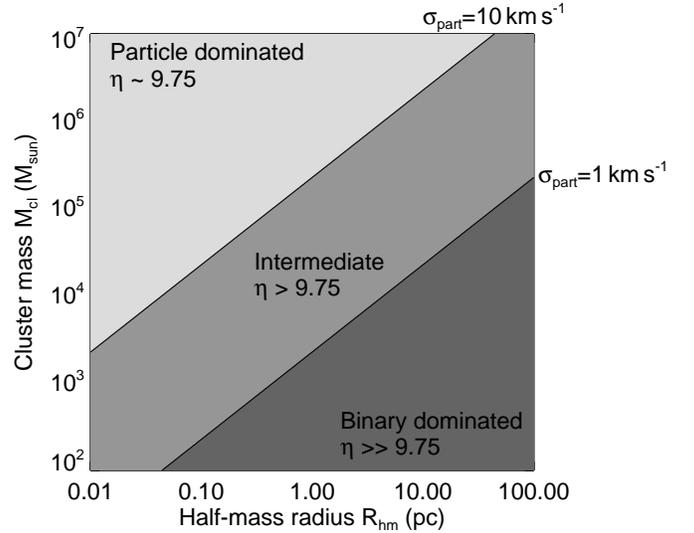}
  \caption{When can the presence of binaries be ignored? The grey-shades in the figure indicate whether the derived dynamical mass is reliable (light) or overestimated (dark). The solid lines indicate models with a centre-of-mass velocity dispersion $\sigmapart=1$~\kms{} (bottom) and 10~\kms{} (top). In the bottom-right and central region, the dynamical mass of the cluster is overestimated by a factor $>2$ and a factor of $1.05-2$, respectively. In the top-left region the dynamical mass overestimation is less than 5\%. Each model has a binary population identical to that of model~R, with a binary fraction of 100\%. For models with a lower binary fraction, the lines move towards the bottom right. 
    \label{figure:effect_R_msun_vs_halfmass} }
\end{figure}


\begin{figure*}[!bt]
  \centering
  \includegraphics[width=\textwidth,height=!]{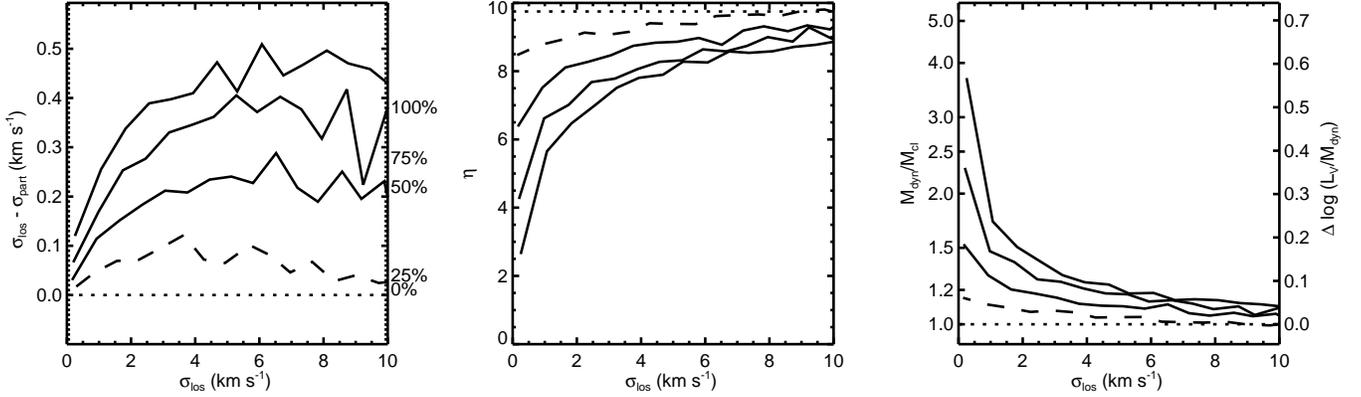}
  \caption{When can binarity be ignored? This figure shows from an observational point of view how the measured velocity dispersion $\sigmalos$ should be interpreted (adopting a binary population as in model~R). The solid curves in each panel represent models with a binary fraction of 25\%, 50\%, 75\% and 100\%, respectively. The dotted curves indicate the results for a cluster with zero binary fraction, and are also representative of the particle motion in each cluster (irrespective of the binary fraction). {\em Left}: the difference between the measured line-of-sight velocity dispersion $\sigmalos$ and the intrinsic centre-of-mass velocity dispersion $\sigmapart$. {\em Middle}: the true value of $\eta$ as derived from the simulations. {\em Right}: the dynamical mass overestimation under the assumption that no binaries are present, i.e., using $\eta=9.75$. On the right axis of this panel we additionally indicate the shift in $\log (L_V/\mdyn)$ introduced by this systematic error. This figure is applicable to all clusters with a Plummer density distribution and a binary population such as in model~R, irrespective of their size, mass, or central density. 
\label{figure:all_vs_sigmalos} }
\end{figure*}


In this section we provide general criteria for when the presence of binaries can or cannot be ignored when determining the dynamical mass of a star cluster. We study clusters with different structural parameters ($\mcl$, $N$, $\halfmass$) and different binary fractions $\binfrac$. We keep all other properties fixed to the values for model~R (see Table~\ref{table:twomodels}). The dependence of $\sigmapart$ on the cluster mass $\mcl$ and the half-mass radius $\halfmass$ is shown in Fig.~\ref{figure:effect_R_msun_vs_halfmass}, where we have adopted a binary fraction of 100\%. The value of $\sigmapart$ as a function of $\mcl$ and $\halfmass$ can be derived from Eq.~(\ref{equation:spitzer}): $\sigmapart \propto \sqrt{\mcl/\halfmass}$. From dark grey to light grey, the colours represent binary-dominated, intermediate, and particle-dominated clusters. The regions are separated by iso-dispersion contours of $\sigmapart=1$~\kms{} and 10~\kms{}, respectively. Clusters with these dispersions have an average density of $\sim 50$~stars\,pc$^{-3}$ and $\sim 5\times 10^7$~stars\,pc$^{-3}$ within their half-mass radius, respectively. For a lower binary fraction, the lines move towards the bottom right of the plot.


Fig.~\ref{figure:all_vs_sigmalos} shows from an observational point of view the results for clusters with a binary population such as for model~R. These are valid for any Plummer-like model, irrespective of mass $\mcl$ and radius $\halfmass$. The horizontal axis in each panel represents the measured line-of-sight velocity dispersion $\sigmalos$. The five curves represent models with a binary fraction of 0\%, 25\%, 50\%, 75\% and 100\%. The left-hand panel shows the contribution of the particle motion to $\sigmalos$. Clearly, the difference $\sigmalos-\sigmapart$ increases with increasing binary fraction. Although this difference appears small, in the range of $0.1-0.5$~\kms{}, its effect on the derivation of the dynamical mass can be large. The middle panel shows the derived value of $\eta$ as a function of $\sigmalos$, for the different binary fractions. For large $\sigmalos$ we find $\eta \approx 9.75$ (dotted curve), indicating that Spitzer's equation is a good approximation. The right-hand panel of Fig.~\ref{figure:all_vs_sigmalos} shows the dynamical mass overestimation $\mdyn/\mcl$. The mass overestimation increases with increasing binary fraction and decreasing $\sigmalos$. The panel shows that if one measures $\sigmalos = 6-8$~\kms{}, the dynamical mass overestimation due to binarity is expected to be $10-30\%$.
%
%
The right-hand panel of Fig.~\ref{figure:all_vs_sigmalos} additionally shows the expected deviation of the datapoints in the $\log (L_V/\mdyn)$ vs. age diagram \citep[see, e.g.,][]{bastian2006,degrijs2007review}, due to binarity. Here, we have used the simple calculation $\Delta \log (L_V/\mdyn) = \log (L_V/\mcl) - \log (L_V/\mdyn) = \log (\mdyn/\mcl)$. Clusters with $\sigmalos \ga 3$~\kms{} have $\Delta \log (L_V/\mdyn) = 0.05 - 0.10$. For clusters with $\sigmalos < 3$~\kms{}, the inconsistency can be significantly larger.

Most clusters in the literature for which the dynamical mass is obtained using Eq.~(\ref{equation:spitzer}) are rather massive, often of order $\mcl=10^{5-7}\msun$. This is mainly because of two reasons: (i) they are bright, and thus easy to detect, and (ii) they have a large velocity dispersion which is easily measured using high-resolution spectroscopy. As the observed clusters are generally massive and have a large $\sigmalos$, their dynamical mass overestimation due to binarity is expected to be mild, of order $5\%$. 


%

\section{Discussion and conclusions} \label{section:conclusions}

The total mass of a distant star cluster is often derived from the virial theorem, using line-of-sight velocity dispersion measurements and half-light radii. This dynamical mass, $\mdyn$, is given by Spitzer's equation, Eq.~(\ref{equation:spitzer}), under the assumption that no binary or multiple systems are present. This assumption is frequently made in the analysis of star clusters, for reasons of simplicity, although it is known that most stars are part of a binary or multiple system. Ignoring this fact may lead to a significant overestimation of the cluster mass. In this paper we have studied the validity of this assumption, and how this affects the dynamical mass determination.

The measured line-of-sight velocity dispersion $\sigmalos$ of an unresolved star cluster can be derived from spectral line analysis. The value $\sigmalos$ represents that of the individual stars. The velocity of these stars is not only determined by the binary centre-of-mass motion $\sigmapart$ in the cluster potential, but also by the orbital motion $\sigmaorb$ of the binary stars. If the assumption is made that no binaries are present, $\sigmalos$ overestimates the motion of the binaries in the cluster potential. Application of Spitzer's equation without taking this effect into account then results in an overestimation of $\mdyn$. 

Whether or not binaries are important depends on (i) the star cluster properties: (ii) the properties of the binary population; and (iii) observational selection effects. In our analysis we therefore distinguish three cases: (a) the particle-dominated case, where binaries can be neglected and Spitzer's equation is approximately valid; (b) the intermediate case, for which $\mdyn$ is overestimated by no more than a factor of two, and (c) the binary-dominated case, where the orbital motion of the binaries dominates $\sigmalos$, and $\mdyn$ is significantly overestimated. Particle-dominated clusters have $\sigmalos \ga 10$~\kms{}, intermediate clusters have $1 \la \sigmalos \la 10$~\kms{}, and binary-dominated clusters have $\sigmalos \la 1$~\kms{}. Depending on the cluster properties, binarity can introduce a shift of $\Delta\log (L_V/\mdyn) = 0.02-0.5$ in the mass-to-light ratio vs. age diagram that is often used to study star cluster evolution. The exact values depend on the properties of the star cluster and its binary population.

The dynamical mass of a star cluster is overestimated if binary systems are present but not properly taken into account in the analysis. Whether or not this overestimation is negligible depends on:

(a) {\em Star cluster properties.}
For a star cluster with a non-zero binary fraction, the structural properties of the cluster are important for the derivation of $\mdyn$. Star clusters with a larger stellar density are least affected by the presence of binaries. These are clusters with a small half-mass radius $\halfmass$, a large number of particles $N=S+B$, or a large total mass $\mcl$. Spitzer's equation assumes that the star cluster is in virial equilibrium, i.e., the cluster has a virial ratio $Q = 0.5$. If this is incorrect, the cluster mass is overestimated by a factor $2Q$. 

(b) {\em Binary population properties.}
Clearly, the dynamical mass of star clusters with a larger binary fraction $\binfrac$ is more affected by the presence of binaries. For a binary-dominated cluster, binarity suddenly becomes important for $\binfrac \ga 50\%$, while for the intermediate case the transition is gradual. Whether a cluster is particle-dominated, of the intermediate case, or binary-dominated, is mainly determined by the ratio $\aeff/\halfmass$, where $\aeff$ is the typical binary semi-major axis. In the binary-dominated case we have $\eta \propto N\aeff\halfmass^{-1}$. Models with a smaller average eccentricity and a larger average mass ratio tend to be associated with a larger dynamical mass overestimation. Variations due to the uncertainty in $f_e(e)$ and $f_q(q)$ are rather small, however: $\Delta (\mdyn/\mcl) \la 5\%$.

(c) {\em Selection effects.}
Observational selection effects are important for the dynamical mass determination. Firstly, the dynamical mass overestimation is more severe if the analysis is more concentrated in the central region of the cluster; this may result in a dynamical mass overestimation of up to 40\%. Secondly, the velocity dispersion is often determined from spectral lines. The flux of these spectral lines is often dominated by a very specific group of stars, e.g., stars of a specific spectral type. If this specific subset of stars is not representative for the cluster was a whole \citep[e.g., more massive stars are often found in close, equal-mass binaries;][]{zinneckeryorke2007}, $\mdyn$ will be biased. In \S~\ref{section:massratio} and~\ref{section:eccentricity} we illustrated that our results differ if we study the velocity dispersion for a different stellar mass range. We find that by selecting stars with $M>1 \msun$ in our example, the mass overestimation increased by 80\% with respect to the unbiased sample: $(\mdyn/\mcl)_{\rm M>1 \msun} \approx 1.8\,(\mdyn/\mcl)_{\mbox{\scriptsize all stars}}$. In a subsequent paper, we will thoroughly study how such selection effects affect the derivation of $\mdyn$. 


We have adopted several other assumptions in our analysis, which require further investigation. For example, we have assumed that the properties of the binary population, $\binfrac$, $M_1$, $q$, $e$, $a$ (or $P$), the orientation, and orbital phase, are uncorrelated. We made this assumption because the purpose of our paper was to illustrate the general effect of binaries on the dynamical mass determination. Observations, however, have indicated that correlations between several of these parameters may be present. The binary fraction, for example, is known to decrease with decreasing primary mass \citep[e.g.,][]{fischermarcy1992,sterzik2004}, and the eccentricity $e$ decreases with decreasing orbital period $P$ due to circularization \citep[e.g.,][]{halbwachs2005}.  In the spectral analysis of a star cluster, one is generally sensitive to only a specific stellar mass range. These often bright or evolved stars may exhibit peculiar properties that are not representative of the cluster as a whole. When interpreting specific observations, one has to specifically characterise the binary population of the sampled subset of stars, before deriving the dynamical mass of the cluster.


Furthermore, we have not included triple or higher-order systems in our simulations, as the properties of these systems are very poorly constrained by observations. The studies of \cite{tokovinin2002} and \cite{correia2006} find that $20-30\%$ of the wide visual binaries have a spectroscopic subsystem \citep[see also][]{huyi2008}. However, our results are unlikely to be affected by this assumption, as stable multiple systems are necessarily hierarchical. The vast majority of triple systems, for example, consists of a primary star, a close spectroscopic companion with semi-major axis $a_1$, and a wide visual companion with semi-major axis $a_2$. In order for the triple system to be stable, we must have $a_1 \ll a_2$ \citep[see][for a detailed study]{mardlingaarseth2001}. For the orbital velocities $v_1$ and $v_2$ of the inner and outer stars we have, to first order approximation, $v_1/v_2 \propto \sqrt{a_2/a_1} \gg 1$. Non-inclusion of the outer component would result in a mild additional overestimation of $\sigmapart$, and thus of $\mdyn$


There are several other causes which may lead to a dynamical mass overestimation. Due to the stochastic and fractal nature of star formation, young star clusters are often found in groups. Confusion may lead to a measured velocity dispersion that is dominated by the systemic velocity difference between two star clusters, rather than the particle or binary motion of the systems in each cluster \citep[e.g.,][]{moll2007}. 


Another bias in the dynamical mass may be caused by mass segregation. The mass segregation can be primordial or dynamical \citep[e.g.,][]{hunter1997,bonnelldavies1998,chen2007,mcmillan2007}. \cite{boily2005}, for example, find that for clusters with a projected density of $\sim 10^4 \msun\,\mbox{pc}^{-2}$ or more, the assumption of $\eta =9.75$ results in a significant dynamical mass underestimation. The latter result is confirmed by \cite{fleck2006}: as heavy stars quickly sink to the centre of the cluster potential, and as these dominate the spectral lines that are used to obtain $\sigmalos$, the dynamical mass may be underestimated by a factor of two. We refer to \cite{fleck2006} for a detailed discussion of the effect of mass segregation on $\eta$ and $\mdyn$.


Although we have made several assumptions in this paper, our main conclusion is robust: the presence of binary systems in a stellar population often results in a significant overestimation of the dynamical mass, if one applies Spitzer's equation without properly taking into account multiplicity.


\begin{acknowledgements}
We wish to thank the referee (Sabine Mengel), Simon Goodwin and Olivier Schnurr for useful comments and valuable suggestions. M.K. was supported by PPARC/STFC under grant number PP/D002036/1. 
\end{acknowledgements}


\bibliographystyle{aa}
\bibliography{8897}

\begin{thebibliography}{64}
\expandafter\ifx\csname natexlab\endcsname\relax\def\natexlab#1{#1}\fi

\bibitem[{{Apai} {et~al.}(2007){Apai}, {Bik}, {Kaper}, {Henning}, \&
  {Zinnecker}}]{apai2007}
{Apai}, D., {Bik}, A., {Kaper}, L., {Henning}, T., \& {Zinnecker}, H. 2007,
  \apj, 655, 484

\bibitem[{{Bahcall} {et~al.}(1985){Bahcall}, {Hut}, \&
  {Tremaine}}]{bahcall1985}
{Bahcall}, J.~N., {Hut}, P., \& {Tremaine}, S. 1985, \apj, 290, 15

\bibitem[{{Bastian} \& {Goodwin}(2006)}]{bastiangoodwin2006}
{Bastian}, N. \& {Goodwin}, S.~P. 2006, \mnras, 369, L9

\bibitem[{{Bastian} {et~al.}(2006){Bastian}, {Saglia}, {Goudfrooij},
  {Kissler-Patig}, {Maraston}, {Schweizer}, \& {Zoccali}}]{bastian2006}
{Bastian}, N., {Saglia}, R.~P., {Goudfrooij}, P., {et~al.} 2006, \aap, 448, 881

\bibitem[{{Boily} {et~al.}(2005){Boily}, {Lan{\c c}on}, {Deiters}, \&
  {Heggie}}]{boily2005}
{Boily}, C.~M., {Lan{\c c}on}, A., {Deiters}, S., \& {Heggie}, D.~C. 2005,
  \apjl, 620, L27

\bibitem[{{Bonnell} \& {Davies}(1998)}]{bonnelldavies1998}
{Bonnell}, I.~A. \& {Davies}, M.~B. 1998, \mnras, 295, 691

\bibitem[{{Bosch} {et~al.}(2001){Bosch}, {Selman}, {Melnick}, \&
  {Terlevich}}]{bosch2001}
{Bosch}, G., {Selman}, F., {Melnick}, J., \& {Terlevich}, R. 2001, \aap, 380,
  137

\bibitem[{{Chanam{\'e}} \& {Gould}(2004)}]{chaname2004}
{Chanam{\'e}}, J. \& {Gould}, A. 2004, \apj, 601, 289

\bibitem[{{Chen} {et~al.}(2007){Chen}, {de Grijs}, \& {Zhao}}]{chen2007}
{Chen}, L., {de Grijs}, R., \& {Zhao}, J.~L. 2007, \aj, 134, 1368

\bibitem[{{Close} {et~al.}(1990){Close}, {Richer}, \& {Crabtree}}]{close1990}
{Close}, L.~M., {Richer}, H.~B., \& {Crabtree}, D.~R. 1990, \aj, 100, 1968

\bibitem[{{Cool} \& {Bolton}(2002)}]{cool2002}
{Cool}, A.~M. \& {Bolton}, A.~S. 2002, in Astronomical Society of the Pacific
  Conference Series, Vol. 263, Stellar Collisions, Mergers and their
  Consequences, ed. M.~M. {Shara}, 163

\bibitem[{{Correia} {et~al.}(2006){Correia}, {Zinnecker}, {Ratzka}, \&
  {Sterzik}}]{correia2006}
{Correia}, S., {Zinnecker}, H., {Ratzka}, T., \& {Sterzik}, M.~F. 2006, \aap,
  459, 909

\bibitem[{{de Grijs} {et~al.}(2008){de Grijs}, {Goodwin}, {Kouwenhoven}, \&
  {Kroupa}}]{degrijs2008lowmass}
{de Grijs}, R., {Goodwin}, S.~P., {Kouwenhoven}, M.~B.~N., \& {Kroupa}, P.
  2008, \mnras, submitted

\bibitem[{{de Grijs} \& {Parmentier}(2007)}]{degrijs2007review}
{de Grijs}, R. \& {Parmentier}, G. 2007, ChJA\&A, 7, 155

\bibitem[{{de Grijs} {et~al.}(2005){de Grijs}, {Wilkinson}, \&
  {Tadhunter}}]{degrijs2005}
{de Grijs}, R., {Wilkinson}, M.~I., \& {Tadhunter}, C.~N. 2005, \mnras, 361,
  311

\bibitem[{{Duquennoy} \& {Mayor}(1991)}]{duquennoy1991}
{Duquennoy}, A. \& {Mayor}, M. 1991, \aap, 248, 485

\bibitem[{{Elmegreen} {et~al.}(2000){Elmegreen}, {Efremov}, {Pudritz}, \&
  {Zinnecker}}]{elmegreen2000}
{Elmegreen}, B.~G., {Efremov}, Y., {Pudritz}, R.~E., \& {Zinnecker}, H. 2000,
  Protostars and Planets IV, 179

\bibitem[{{Fischer} \& {Marcy}(1992)}]{fischermarcy1992}
{Fischer}, D.~A. \& {Marcy}, G.~W. 1992, \apj, 396, 178

\bibitem[{{Fleck} {et~al.}(2006){Fleck}, {Boily}, {Lan{\c c}on}, \&
  {Deiters}}]{fleck2006}
{Fleck}, J.-J., {Boily}, C.~M., {Lan{\c c}on}, A., \& {Deiters}, S. 2006,
  \mnras, 369, 1392

\bibitem[{{Goodwin} \& {Bastian}(2006)}]{goodwinbastian2006}
{Goodwin}, S.~P. \& {Bastian}, N. 2006, \mnras, 373, 752

\bibitem[{{Goodwin} \& {Kroupa}(2005)}]{goodwinkroupa2005}
{Goodwin}, S.~P. \& {Kroupa}, P. 2005, \aap, 439, 565

\bibitem[{{Halbwachs} {et~al.}(2005){Halbwachs}, {Mayor}, \&
  {Udry}}]{halbwachs2005}
{Halbwachs}, J.~L., {Mayor}, M., \& {Udry}, S. 2005, \aap, 431, 1129

\bibitem[{{Heggie} \& {Hut}(2003)}]{heggiehut}
{Heggie}, D. \& {Hut}, P. 2003, {The Gravitational Million-Body Problem: A
  Multidisciplinary Approach to Star Cluster Dynamics} (Cambridge University
  Press)

\bibitem[{{Heggie}(1975)}]{heggie1975}
{Heggie}, D.~C. 1975, \mnras, 173, 729

\bibitem[{{Hu} {et~al.}(2008){Hu}, {Deng}, {de Grijs}, {Goodwin}, \&
  {Qiang}}]{huyi2008}
{Hu}, Y., {Deng}, L., {de Grijs}, R., {Goodwin}, S.~P., \& {Qiang}, L. 2008,
  MNRAS, submitted

\bibitem[{{Hunter} {et~al.}(1997){Hunter}, {Light}, {Holtzman}, {Lynds},
  {O'Neil}, \& {Grillmair}}]{hunter1997}
{Hunter}, D.~A., {Light}, R.~M., {Holtzman}, J.~A., {et~al.} 1997, \apj, 478,
  124

\bibitem[{{King}(1962)}]{king1962}
{King}, I. 1962, \aj, 67, 471

\bibitem[{{King}(1966)}]{king1966}
{King}, I.~R. 1966, \aj, 71, 64

\bibitem[{{Kobulnicky} \& {Fryer}(2007)}]{kobulnicky2007}
{Kobulnicky}, H.~A. \& {Fryer}, C.~L. 2007, \apj, 670, 747

\bibitem[{{Kouwenhoven} {et~al.}(2007){Kouwenhoven}, {Brown}, {Portegies
  Zwart}, \& {Kaper}}]{kouwenhoven_recovery}
{Kouwenhoven}, M.~B.~N., {Brown}, A.~G.~A., {Portegies Zwart}, S.~F., \&
  {Kaper}, L. 2007, \aap, 474, 77

\bibitem[{{Kouwenhoven} {et~al.}(2005){Kouwenhoven}, {Brown}, {Zinnecker},
  {Kaper}, \& {Portegies Zwart}}]{kouwenhoven_adonis}
{Kouwenhoven}, M.~B.~N., {Brown}, A.~G.~A., {Zinnecker}, H., {Kaper}, L., \&
  {Portegies Zwart}, S.~F. 2005, \aap, 430, 137

\bibitem[{{Kroupa}(2001)}]{kroupa2001}
{Kroupa}, P. 2001, \mnras, 322, 231

\bibitem[{{Kroupa} \& {Boily}(2002)}]{kroupaboily2002}
{Kroupa}, P. \& {Boily}, C.~M. 2002, \mnras, 336, 1188

\bibitem[{{Larsen} {et~al.}(2004){Larsen}, {Brodie}, \& {Hunter}}]{larsen2004}
{Larsen}, S.~S., {Brodie}, J.~P., \& {Hunter}, D.~A. 2004, \aj, 128, 2295

\bibitem[{{Larsen} {et~al.}(2007){Larsen}, {Origlia}, {Brodie}, \&
  {Gallagher}}]{larsen2007}
{Larsen}, S.~S., {Origlia}, L., {Brodie}, J.~P., \& {Gallagher}, III, J.~S.
  2007, MNRAS, accepted (ArXiv:0710.0547)

\bibitem[{{Lucy}(2006)}]{lucy2006}
{Lucy}, L.~B. 2006, \aap, 457, 629

\bibitem[{{Ma} {et~al.}(2006){Ma}, {de Grijs}, {Yang}, {Zhou}, {Chen}, {Jiang},
  {Wu}, \& {Wu}}]{ma2006}
{Ma}, J., {de Grijs}, R., {Yang}, Y., {et~al.} 2006, \mnras, 368, 1443

\bibitem[{{Mandushev} {et~al.}(1991){Mandushev}, {Spasova}, \&
  {Staneva}}]{mandushev1991}
{Mandushev}, G., {Spasova}, N., \& {Staneva}, A. 1991, \aap, 252, 94

\bibitem[{{Maraston} {et~al.}(2004){Maraston}, {Bastian}, {Saglia},
  {Kissler-Patig}, {Schweizer}, \& {Goudfrooij}}]{maraston2004}
{Maraston}, C., {Bastian}, N., {Saglia}, R.~P., {et~al.} 2004, \aap, 416, 467

\bibitem[{{Mardling} \& {Aarseth}(2001)}]{mardlingaarseth2001}
{Mardling}, R.~A. \& {Aarseth}, S.~J. 2001, \mnras, 321, 398

\bibitem[{{Mason} {et~al.}(1998){Mason}, {Gies}, {Hartkopf}, {Bagnuolo}, {ten
  Brummelaar}, \& {McAlister}}]{mason1998}
{Mason}, B.~D., {Gies}, D.~R., {Hartkopf}, W.~I., {et~al.} 1998, \aj, 115, 821

\bibitem[{{Mathieu}(1994)}]{mathieu1994}
{Mathieu}, R.~D. 1994, \araa, 32, 465

\bibitem[{{McMillan} {et~al.}(2007){McMillan}, {Vesperini}, \& {Portegies
  Zwart}}]{mcmillan2007}
{McMillan}, S.~L.~W., {Vesperini}, E., \& {Portegies Zwart}, S.~F. 2007, \apjl,
  655, L45

\bibitem[{{Mengel} {et~al.}(2005){Mengel}, {Lehnert}, {Thatte}, \&
  {Genzel}}]{mengel2005}
{Mengel}, S., {Lehnert}, M.~D., {Thatte}, N., \& {Genzel}, R. 2005, \aap, 443,
  41

\bibitem[{{Meylan} \& {Heggie}(1997)}]{meylan1997}
{Meylan}, G. \& {Heggie}, D.~C. 1997, \aapr, 8, 1

\bibitem[{{Moll} {et~al.}(2007){Moll}, {Mengel}, {de Grijs}, {Smith}, \&
  {Crowther}}]{moll2007}
{Moll}, S.~L., {Mengel}, S., {de Grijs}, R., {Smith}, L.~J., \& {Crowther},
  P.~A. 2007, \mnras, 1049

\bibitem[{{\"{O}pik}(1924)}]{opik1924}
{\"{O}pik}, E. 1924, Tartu Obs. Publ., 25, No. 6

\bibitem[{{Pinsonneault} \& {Stanek}(2006)}]{pinsonneault2006}
{Pinsonneault}, M.~H. \& {Stanek}, K.~Z. 2006, \apjl, 639, L67

\bibitem[{{Plummer}(1911)}]{plummer1911}
{Plummer}, H.~C. 1911, \mnras, 71, 460

\bibitem[{{Portegies Zwart} {et~al.}(2001){Portegies Zwart}, {McMillan}, {Hut},
  \& {Makino}}]{ecology4}
{Portegies Zwart}, S.~F., {McMillan}, S.~L.~W., {Hut}, P., \& {Makino}, J.
  2001, \mnras, 321, 199

\bibitem[{{Poveda} \& {Allen}(2004)}]{poveda2004}
{Poveda}, A. \& {Allen}, C. 2004, in Revista Mexicana de Astronomia y
  Astrofisica Conference Series, ed. C.~{Allen} \& C.~{Scarfe}, 49

\bibitem[{{Poveda} {et~al.}(2007){Poveda}, {Allen}, \&
  {Hern{\'a}ndez-Alc{\'a}ntara}}]{poveda2007}
{Poveda}, A., {Allen}, C., \& {Hern{\'a}ndez-Alc{\'a}ntara}, A. 2007, in IAU
  Symposium, Vol. 240, IAU Symposium, 417

\bibitem[{{Reijns} {et~al.}(2006){Reijns}, {Seitzer}, {Arnold}, {Freeman},
  {Ingerson}, {van den Bosch}, {van de Ven}, \& {de Zeeuw}}]{reijns2006}
{Reijns}, R.~A., {Seitzer}, P., {Arnold}, R., {et~al.} 2006, \aap, 445, 503

\bibitem[{{Smith} \& {Gallagher}(2001)}]{smith2001}
{Smith}, L.~J. \& {Gallagher}, J.~S. 2001, \mnras, 326, 1027

\bibitem[{{S{\"o}derhjelm}(2007)}]{soderhjelm2007}
{S{\"o}derhjelm}, S. 2007, \aap, 463, 683

\bibitem[{{Sollima} {et~al.}(2007){Sollima}, {Beccari}, {Ferraro}, {Fusi
  Pecci}, \& {Sarajedini}}]{sollima2007}
{Sollima}, A., {Beccari}, G., {Ferraro}, F.~R., {Fusi Pecci}, F., \&
  {Sarajedini}, A. 2007, \mnras, 380, 781

\bibitem[{{Spitzer}(1987)}]{spitzer1987}
{Spitzer}, L. 1987, {Dynamical evolution of globular clusters} (Princeton
  University Press)

\bibitem[{{Sterzik} \& {Durisen}(2004)}]{sterzik2004}
{Sterzik}, M.~F. \& {Durisen}, R.~H. 2004, in Revista Mexicana de Astronomia y
  Astrofisica Conference Series, Vol.~21, The Environment and Evolution of
  Double and Multiple Stars, ed. C.~{Allen} \& C.~{Scarfe}, 58

\bibitem[{{Stolte} {et~al.}(2007){Stolte}, {Ghez}, {Morris}, {Lu}, {Brandner},
  \& {Matthews}}]{stolte2007}
{Stolte}, A., {Ghez}, A.~M., {Morris}, M., {et~al.} 2007, ApJ, submitted
  (ArXiv:0706.4133)

\bibitem[{{Tokovinin} \& {Smekhov}(2002)}]{tokovinin2002}
{Tokovinin}, A.~A. \& {Smekhov}, M.~G. 2002, \aap, 382, 118

\bibitem[{{van Albada}(1968)}]{vanalbada1968}
{van Albada}, T.~S. 1968, \bain, 20, 47

\bibitem[{{van Leeuwen} {et~al.}(2000){van Leeuwen}, {Le Poole}, {Reijns},
  {Freeman}, \& {de Zeeuw}}]{vanleeuwen2000}
{van Leeuwen}, F., {Le Poole}, R.~S., {Reijns}, R.~A., {Freeman}, K.~C., \& {de
  Zeeuw}, P.~T. 2000, \aap, 360, 472

\bibitem[{{Vereshchagin} {et~al.}(1988){Vereshchagin}, {Tutukov}, {Iungelson},
  {Kraicheva}, \& {Popova}}]{vereshchagin1988}
{Vereshchagin}, S., {Tutukov}, A., {Iungelson}, L., {Kraicheva}, Z., \&
  {Popova}, E. 1988, \apss, 142, 245

\bibitem[{{Zinnecker} \& {Yorke}(2007)}]{zinneckeryorke2007}
{Zinnecker}, H. \& {Yorke}, H.~W. 2007, \araa, 45, 481

\end{thebibliography}




\end{document}